\documentclass[12pt]{article}
\usepackage{epsfig}
\usepackage{psfrag}
\usepackage{latexsym}
\usepackage[DIV13]{typearea}
\usepackage{amsmath}
\usepackage{amssymb}
\usepackage{amsfonts}  
\usepackage{cite}
\usepackage{bbold}
\usepackage[footnotesize]{caption2}
\usepackage{graphicx}
\usepackage[center,footnotesize,hang]{subfigure}
\usepackage{url}
\usepackage{color}

\newcommand{\Uu}{\text{U(1)}}

\def\gs{\mathrel{
   \rlap{\raise 0.511ex \hbox{$>$}}{\lower 0.511ex \hbox{$\sim$}}}}
\def\ls{\mathrel{
   \rlap{\raise 0.511ex \hbox{$<$}}{\lower 0.511ex \hbox{$\sim$}}}}

\newcommand{\ba}{\begin{array}{c}}
\newcommand{\baz}{\begin{array}{cc}}
\newcommand{\bad}{\begin{array}{ccc}}
\newcommand{\ea}{\end{array}}

\newcommand{\be}{\beta}

\newcommand{\La}{\Lambda}

\def\beq{\begin{equation}}
\def\eeq{\end{equation}}
\def\bea{\begin{eqnarray}}
\def\eea{\end{eqnarray}}
\def\bet{\begin{tabular}}
\def\eet{\end{tabular}}
\def\bes{\begin{subequations}\bea}
\def\ees{\eea\end{subequations}}

\def\be{\begin{equation}}
\def\ee{\end{equation}}
\def\bc{\begin{center}}
\def\ec{\end{center}}
\def\bea{\begin{eqnarray}}
\def\eea{\end{eqnarray}}
\def\dd{\displaystyle}
\def\nn{\nonumber}

\def\ov{\overline}

\catcode`@=11
\def\marginnote#1{}
\newcount\hour
\newcount\minute
\newtoks\amorpm
\hour=\time\divide\hour by60
\minute=\time{\multiply\hour by60 \global\advance\minute by-\hour}
\edef\standardtime{{\ifnum\hour<12 \global\amorpm={am}%
        \else\global\amorpm={pm}\advance\hour by-12 \fi
        \ifnum\hour=0 \hour=12 \fi
        \number\hour:\ifnum\minute<10 0\fi\number\minute\the\amorpm}}
\edef\militarytime{\number\hour:\ifnum\minute<10 0\fi\number\minute}
\def\draftlabel#1{{\@bsphack\if@filesw {\let\thepage\relax
   \xdef\@gtempa{\write\@auxout{\string
      \newlabel{#1}{{\@currentlabel}{\thepage}}}}}\@gtempa
   \if@nobreak \ifvmode\nobreak\fi\fi\fi\@esphack}
        \gdef\@eqnlabel{#1}}
\def\@eqnlabel{}
\def\@vacuum{}
\def\draftmarginnote#1{\marginpar{\raggedright\scriptsize\tt#1}}
\def\draft{\oddsidemargin 0.0truein
        \def\@oddfoot{\sl preliminary draft \hfil
        \rm\thepage\hfil\sl\today\quad\militarytime}
        \let\@evenfoot\@oddfoot \overfullrule 3pt
        \let\label=\draftlabel
        \let\marginnote=\draftmarginnote
   \def\@eqnnum{(\theequation)\rlap{\kern\marginparsep\tt\@eqnlabel}%
\global\let\@eqnlabel\@vacuum}  }
\catcode`@=12

\begin{document}
\begin{titlepage}
\vspace*{-1cm}
\phantom{hep-ph/***}
\hfill{DFPF-09/TH/18}\hspace{3.25in}\hfill{SISSA  66/2009/EP}\\
\vskip 2.5cm
\begin{center}
\mathversion{bold}
{\Large\bf Vacuum Alignment in SUSY $A_4$ Models}
\mathversion{normal}
\end{center}
\vskip 0.5  cm
\begin{center}
{\large Ferruccio Feruglio}~$^{a)}$\footnote{e-mail address: feruglio@pd.infn.it},
{\large Claudia Hagedorn}~$^{b)}$\footnote{e-mail address: hagedorn@sissa.it}
and {\large Luca Merlo}~$^{a)}$\footnote{e-mail address: merlo@pd.infn.it}
\\
\vskip .2cm
$^{a)}$~Dipartimento di Fisica `G.~Galilei', Universit\`a di Padova
\\
INFN, Sezione di Padova, Via Marzolo~8, I-35131 Padua, Italy
\\
\vskip .1cm
$^{b)}$~
SISSA, Scuola Internazionale Superiore di Studi Avanzati\\
Via Beirut 2-4, I-34014 Trieste, Italy\\
and\\ INFN, Sezione di Trieste, Italy
\end{center}
\vskip 0.7cm
\begin{abstract}
\noindent
In this note we discuss the vacuum alignment in globally supersymmetric models
with spontaneously broken flavour symmetries in the presence of generic
soft supersymmetry (SUSY) breaking terms. We show that the inclusion of these
soft SUSY breaking terms can give rise to non-vanishing vacuum expectation values (VEVs)
for the auxiliary components of the flavon fields.
These non-zero VEVs can have an important impact on the phenomenology of this class of models, since they can
induce an additional flavour violating contribution to the sfermion soft mass matrix of right-left (RL) type.
We carry out an explicit computation in a class of globally SUSY $A_4$ models
predicting tri-bimaximal mixing in the lepton sector. The flavour symmetry breaking sector is described in terms of
flavon and driving supermultiplets. 
We find non-vanishing VEVs for the
auxiliary components of the flavon fields and for the scalar components of the driving fields which are of
order $m_{SUSY} \times \langle\varphi\rangle$ and $m_{SUSY}$, respectively. 
Thereby, $m_{SUSY}$ is the generic soft SUSY breaking scale which is expected to be around 1 TeV and
$\langle\varphi\rangle$ is the VEV of scalar components of the flavon fields.
Another effect of these VEVs can be the generation of a $\mu$ term. 
\end{abstract}
\end{titlepage}
\setcounter{footnote}{0}
\vskip2truecm
\section{Introduction}
A well-known problem of SUSY extensions of the Standard Model (SM) with superparticles at the TeV scale is the presence of
new sources of flavour violation, see e.g. \cite{Martin}. They are due to the couplings between SUSY particles and ordinary particles
that, for generic soft SUSY breaking terms, are incompatible with the present limits on rare flavour-changing transitions.
It is reasonable to expect that in the presence of a flavour symmetry, helpful to reproduce the observed pattern of
fermion masses and mixing angles, the soft SUSY breaking terms are more constrained and the above problem is alleviated.
We consider a globally SUSY framework where the soft SUSY breaking terms are already present at the scale relevant to flavour dynamics,
so that their boundary conditions at that scale are dictated by the flavour symmetry.

An important aspect of the problem is the relative alignment in flavour space of fermion and sfermion mass terms.
Focusing on the lepton sector, in models invariant under a flavour symmetry group, the lepton masses are
described by a superpotential of the type
\be
\label{superp}
w=e^c_\alpha H_d {\cal Y}_{\alpha\beta}(\varphi) l_\beta+w_d(\varphi)+...
\ee
where $\varphi_i$ denotes the set of chiral superfields neutral under gauge interactions, the flavons, whose scalar components break the flavour symmetry through their VEVs.
The analytic functions ${\cal Y}_{\alpha\beta}(\varphi)$ and $w_d(\varphi)$
depend only on the supermultiplets $\varphi_i$ and they generally admit an expansion in inverse powers of some ultraviolet cutoff $\Lambda_f$
representing the flavour scale, such that $\langle\varphi_i\rangle/\Lambda_f$ are small parameters and one can truncate the expansion
after the first few terms. In the SUSY limit the term $w_d(\varphi)$ determines the scalar potential of the flavons and is responsible for the
breaking of the flavour symmetry. The lepton mass matrix is proportional to ${\cal Y}_{\alpha\beta}(\langle\varphi\rangle)$, while
slepton masses receive contributions of different types. In this note we are interested in the contribution
arising in the scalar potential after eliminating the auxiliary fields of the flavon supermultiplets. Assuming a canonical K\"ahler potential we get
\be
V=\tilde{e}^c_\alpha  H_d \left\langle\dd\frac{\partial {\cal Y}_{\alpha\beta}}{\partial \varphi_i} \ov{\dd\frac{\partial w_d}{\partial \varphi_i}}\right\rangle\, \tilde{l}_\beta+ \mathrm{h.c.}+...
\label{v}
\ee
Dots stand for additional contributions, not relevant in this context.
In the SUSY limit of course the auxiliary components of the flavons vanish at the minimum, that is $\langle \partial w_d/\partial \varphi_i \rangle=0$ and the above contribution vanishes, but including SUSY breaking effects in general we expect $\langle \partial w_d/\partial \varphi_i \rangle\ne 0$
and of order $m_{SUSY} \times \langle \varphi_i \rangle$. Thus it is important to establish whether the combinations
\be
\left\langle\dd\frac{\partial {\cal Y}_{\alpha\beta}}{\partial \varphi_i} \ov{\dd\frac{\partial w_d}{\partial \varphi_i}}\right\rangle
\ee
and
\be
{\cal Y}_{\alpha\beta}(\langle\varphi\rangle)
\ee
can be simultaneously diagonalized (in flavour space) or not, since a misalignment would represent a source of flavour violation.
If the expansion of ${\cal Y}_{\alpha\beta}(\varphi)$ is linear in $\varphi_i$ at the leading order (LO)
\be
{\cal Y}_{\alpha\beta}(\varphi)={\cal Y}_{\alpha\beta}^{(1)i}~\dd\frac{\varphi_i}{\Lambda_f}+...
\label{Yabexp}
\ee
where ${\cal Y}_{\alpha\beta}^{(1)i}$ are constants, a necessary condition for the alignment of the above combinations is
\be
\left\langle \dd\ov{\frac{\partial w_d}{\partial \varphi_i}} \right\rangle=\alpha \langle\varphi_i\rangle~~~,
\label{align}
\ee
where $\alpha$ is a constant. Note that $\alpha$ has to be the same for all (irreducible) flavon multiplets which couple to 
charged leptons at this order. Whether this condition is sufficient to eliminate all relevant sources of flavour violation 
associated to the contribution given in eq. (\ref{v}), depends in general also on the structure of the subleading terms represented
by dots in the expansion in eq. (\ref{Yabexp}). Another possibility to render the effect of the contribution in eq. (\ref{v}) phenomenologically harmless would be to suppress the size of $\langle\partial w_d/\partial \varphi_i\rangle$ below its generic value of $m_{SUSY}
\times \langle \varphi_i \rangle$. As it has been shown in \cite{dynamicsupp} this is indeed possible 
in the context of supergravity. In general,  
the VEVs of the auxiliary components of flavon fields are expected to be of order $m_{3/2} \times \langle \varphi_i\rangle$ \cite{RV}, $m_{3/2}$ being the
gravitino mass and $m_{3/2} \sim m_{SUSY}$, due to the contribution to the $F$-term proportional to the superpotential $w$
\footnote{For further examples see \cite{Ftermsnewer}.}. However, as discussed in \cite{dynamicsupp} this 
contribution can be canceled against the global SUSY contribution to the $F$-terms so that the VEVs of the auxiliary components
are $\ll m_{3/2} \times \langle \varphi_i \rangle$. We will comment on this possibility in Section 4.

In this note we analyze $\langle\partial w_d/\partial \varphi_i\rangle$ in the specific case of a globally SUSY model with $A_4$
flavour symmetry  \cite{af2,a4models} in which tri-bimaximal mixing \cite{TB} can be successfully generated in the lepton sector.
We show explicitly through minimization of the flavon potential including generic soft SUSY breaking terms that
the auxiliary components of the flavons acquire in general non-vanishing VEVs.
From the explicit expressions of these VEVs, we show that
there is a special case in which they vanish, corresponding to universal soft SUSY breaking terms in the flavon potential.
Furthermore, we show that the possibility of completely aligned VEVs of flavons and their auxiliary components, eq. (\ref{align}), 
has to be considered as a fine-tuning.
We comment on the impact of this effect on lepton flavour violating decays in the concluding section, but we leave
a detailed discussion for a separate work \cite{fhlm09}.
Additionally, we confirm the result for the VEVs of the flavons at the LO and the next-to-leading order (NLO) of \cite{af2}.
We note that we perform the calculation in the limit of canonical kinetic terms, although these
are in general non-canonical. We will comment on this assumption below.
Furthermore, we comment on the introduction of a $\mu$ term and also on the mass spectrum in the flavour symmetry breaking sector at
the LO in the SUSY limit, which indicates the presence of two real flat directions which give rise to the undetermined
complex parameter in the flavon VEVs.

\section{The SUSY limit}
We consider a class of SUSY models invariant under a discrete flavour symmetry group, $A_4$.
In the simplest case the gauge group is the SM one.
Crucial ingredients of this type of models are the following: $(a)$ additional degrees
of freedom, flavons and driving fields, which are responsible for the breaking of the
flavour symmetry and which do not transform under the gauge group and $(b)$ additional symmetries
apart from $A_4$ which are necessary for achieving the vacuum alignment. In this note
we assume that $A_4$ is accompanied by a cyclic symmetry $Z_3$, necessary to separate the
charged lepton and the neutrino sector, a continuous $R$ symmetry $\Uu_R$, simplifying the construction of the
scalar potential and a Froggatt-Nielsen symmetry $\Uu_{FN}$ \cite{fn} giving rise to the
charged lepton mass hierarchy and not relevant for the present discussion.
The following flavons and driving fields are assumed in this model: a triplet
$\varphi_T$ giving masses to charged leptons at the LO, two singlets $\xi$ and $\tilde\xi$
and another triplet $\varphi_S$ leading to neutrino masses and driving fields $\varphi^T_0$,
$\varphi^S_0$ and $\xi_0$. These are collected in Table 1.
Flavons have a vanishing charge under $\Uu_R$, whereas driving fields
are assigned the charge $+2$. Through this the superpotential is linear in the driving fields.
The model also includes two electroweak doublets, $H_{u,d}$, responsible for electroweak
symmetry breaking. In the minimization of the scalar potential we 
work in the limit $\langle H_{u,d}\rangle=0$.

In the SUSY limit the $F$-terms of all fields are required to vanish and from the
conditions
\beq
\left\langle\frac{\partial w}{\partial \varphi^T_0}\right\rangle  =0 \; , \;\;
\left\langle\frac{\partial w}{\partial \varphi^S_0} \right\rangle =0 \;\;\; \mbox{and} \;\;\;
\left\langle \frac{\partial w}{\partial \xi_0} \right\rangle =0
\eeq
we can derive
\bea
\frac{\langle\varphi_T\rangle}{\Lambda_f}&=&(\epsilon,0,0)+(c' \epsilon^2,c \epsilon^2,c \epsilon^2) \; , \;\;
\frac{\langle\varphi_S\rangle}{\Lambda_f}= c_b(\epsilon,\epsilon,\epsilon)+(c_1 \epsilon^2, c_2 \epsilon^2, c_3 \epsilon^2) \; , \;\; \nonumber\\
\frac{\langle\xi\rangle}{\Lambda_f}&=&c_a \epsilon \;\;\; \mbox{and} \;\;\;
\frac{\langle\tilde\xi\rangle}{\Lambda_f}=c_c \epsilon^2
\label{flavonVEVs}
\eea
where $c'$, $c$, $c_{a,b,c}$ and $c_i$ are complex numbers with absolute value of order one,
depending on one undetermined parameter.
The undetermined parameter indicates that there is a flat direction in the subspace ($\varphi_S$, $\xi$).
The parameter $\epsilon$ is defined as the VEV of the first component of the triplet $\varphi_T$,
in units of the cutoff scale $\Lambda_f$. In this model $\epsilon$ is a small parameter,
which controls the expansion in inverse powers of $\Lambda_f$.
Its typical size is $0.007 \lesssim \epsilon \lesssim 0.05$.
We have displayed the result at the NLO, up to and including $\epsilon^2$ terms.
At the same time we deduce from the vanishing of the $F$-terms associated to the flavons,
\beq
\left\langle\frac{\partial w}{\partial \varphi_T}\right\rangle  =0 \; , \;\;
\left\langle\frac{\partial w}{\partial \varphi_S} \right\rangle =0 \; , \;\;\;
\left\langle \frac{\partial w}{\partial \xi} \right\rangle =0\;\;\; \mbox{and} \;\;\;
\left\langle \frac{\partial w}{\partial \tilde{\xi}} \right\rangle =0 \; ,
\eeq
that the VEVs of the driving fields vanish in the SUSY limit, to all orders in the expansion parameter $\epsilon$.
This result has been discussed in detail in \cite{af2} and we recover it as a byproduct of our computation. We just recall that the special pattern of VEVs in eq. (\ref{flavonVEVs})
is the crucial ingredient to reproduce the tri-bimaximal mixing in this class of models. The explicit expressions of these VEVs are shown in the Appendix.
\begin{table}[h!]
\begin{center}
\begin{tabular}{|c||c|c|c|c||c|c|c||c|}
\hline
{\tt Field}& $\varphi_T$ & $\varphi_S$ & $\xi$ & $\tilde{\xi}$ & $\varphi_0^T$ & $\varphi_0^S$ & $\xi_0$ & $H_{u,d}$\\
\hline
$A_4$ & $3$ & $3$ & $1$ & $1$ & $3$ & $3$ & $1$ & $1$\\
\hline
$Z_3$ & $1$ & $\omega$ & $\omega$ & $\omega$ & $1$ & $\omega$ & $\omega$ & $1$\\
\hline
$\Uu_R$ & $0$ & $0$ & $0$ & $0$ & $2$ & $2$ & $2$ & $0$\\
\hline
\end{tabular}
\end{center}
\begin{center}
\normalsize
\begin{minipage}[t]{15cm}
\caption[Flavons and Driving fields in the model]{Supermultiplets
of the model given in \cite{af2}, relevant for the breaking of the flavour symmetry $A_4 \times Z_3$,
the electroweak doublets $H_u$, $H_d$ and their transformation properties under the symmetries
$A_4 \times Z_3 \times \Uu_R$.
\label{table}}
\end{minipage}
\end{center}
\end{table}

The presence of an undetermined VEV in the flavon sector is associated to the existence of a complex flat direction in the flavon potential.
This can be checked by analyzing
the mass spectrum of flavons and driving fields in the SUSY limit. In doing so we find that the massive modes have masses either proportional
to the mass parameter of the superpotential $w_d$ or to the undetermined VEV.
This mass spectrum receives corrections of several types: from the NLO terms in the scalar potential, from deviations of the K\"ahler potential
from the assumed canonical form, from SUSY breaking effects and from radiative corrections. The two last classes of corrections can be important to give mass to the modes associated to the flat direction as well as to stabilize the undetermined VEV around a value
of order $\epsilon~ \Lambda_f$.

%
\section{Generic soft SUSY breaking terms}
We proceed to include generic soft SUSY breaking terms, which originate from another sector of the theory, 
 completely neutral under the action of the gauge
group and under $A_4\times Z_3\times \Uu_{FN}$. 
However, without referring to a specific SUSY breaking mechanism this sector remains undetermined and thus also no further
information of the form of the soft SUSY breaking terms can be used.
The search for field configurations that minimize the energy density
by simultaneously varying both the field content of the SUSY breaking sector and the flavon fields is beyond the scope of this work,
 but we present a schematic discussion of this point in Section 4, where we embed our model into a supergravity framework.
In our model we mimic the new sector through a set of generic soft SUSY breaking terms
obtained by promoting the coupling constants of the theory to constant superfields
with non-vanishing auxiliary components \cite{LutyReview}. For instance a superpotential coupling constant
$\mathbf{g}$ is expanded as  $g + \mathbb{g}~ m_{SUSY} \theta^2$. 
In our analysis we work under the assumption that the dynamics of the flavon sector
do not appreciably affect that of the SUSY breaking sector, so that at the scale $\Lambda_f$ and below
we can separately discuss the minimization of the scalar potential with respect to the
flavons and the driving fields, in the presence of fixed soft SUSY breaking terms for them.

We discuss the K\"ahler potential of the flavon and driving fields. We make the
assumption that it is canonical even though the symmetries of the model would
allow subleading corrections to the canonical form. We show that, even in the absence
of these corrections, in general the alignment in eq. (\ref{align}) is not realized.
We do not expect that the introduction of a new set of parameters related to the non-canonical
part of the K\"ahler potential could restore the alignment, unless a fine-tuning is enforced.
The K\"ahler potential represents also a possible source of soft SUSY breaking terms for the flavons and the
driving fields, once we promote its parameters to constant superfields. In the $D$-term for the flavons and the driving
fields, collectively denoted as $\varphi_i$
\be
\int d^2\theta d^2\ov{\theta}~ \ov{\varphi}_i Z_{ij}\varphi_j
\ee
we regard the matrix $Z_{ij}$ as a constant superfield
\be
Z_{ij}=\delta_{ij}+\Gamma_{ij}m_{SUSY} \theta^2+\ov{\Gamma}_{ji} m_{SUSY}\ov{\theta}^2+C_{ij}m_{SUSY}^2\theta^2 \ov{\theta}^2
\ee
where the first term gives rise to the kinetic terms, while the remaining ones generate soft SUSY breaking terms.
It is not restrictive to set the matrix $\Gamma_{ij}$ to zero in the
above decomposition. Indeed, as can be shown, after eliminating the auxiliary fields the effect of a non-vanishing $\Gamma_{ij}$
can be absorbed by redefining the matrix $C_{ij}$ and the parameters describing the soft SUSY
breaking terms originating from the superpotential $w_d(\varphi)$. By setting $\Gamma_{ij}$ to zero
we are left with the contribution of $C_{ij}$ which is proportional to $m_{SUSY}^2$.
When calculating the VEVs of flavons and driving fields including soft SUSY breaking terms into the flavon potential,
we perform a double expansion in the small symmetry breaking parameter $\epsilon$ and in the soft SUSY breaking mass $m_{SUSY}$
which is assumed to be much smaller than the cutoff scale of the theory $\La_f$ and the VEVs of the flavons.
As we see below, the quantities $\langle\partial w/\partial \varphi_i\rangle$
we are interested in are proportional to $m_{SUSY} \times \epsilon \, \Lambda_f$, at the LO in our expansion.
One can show that the SUSY breaking terms coming from the K\"ahler potential contribute to
$\langle\partial w/\partial \varphi_i\rangle$ with terms at the order $m_{SUSY}^2\ll m_{SUSY}\times \epsilon \, \Lambda_f$
and therefore they can be safely neglected in our analysis. Thus we take $Z_{ij}=\delta_{ij}$ in the following.

Under the above assumptions, the relevant object of our computation is the superpotential of the theory
and its dependence upon the flavons and the driving fields.
From this we subsequently derive the soft SUSY breaking terms by promoting the coupling constants in the superpotential
to superfields with constant $\theta^2$ components following \cite{LutyReview}.
The superpotential $w_d$ of flavons and driving fields at the NLO level, generated by including terms which are suppressed by one power of the cutoff scale $\La_f$,
has already been calculated and discussed in detail \cite{af2}. In order to establish
our notation we repeat the results found in \cite{af2}.
The superpotential can be expanded in the parameter $\epsilon$
\beq
w_d = w_d^{(0)} + w_d^{(1)} + ...
\eeq
At the LO the superpotential $w_d$ is of the form
\footnote{We change the notation slightly compared to the original version found in \cite{af2} by renaming $g$ into $g_0$.}
\bea
w_d^{(0)}&=&\mathbf{M} (\varphi_0^T \varphi_T)+ \mathbf{g_0} (\varphi_0^T \varphi_T\varphi_T)\nn\\
&+&\mathbf{g_1} (\varphi_0^S \varphi_S\varphi_S)+
\mathbf{g_2} \tilde{\xi} (\varphi_0^S \varphi_S)+
\mathbf{g_3} \xi_0 (\varphi_S\varphi_S)+
\mathbf{g_4} \xi_0 \xi^2+
\mathbf{g_5} \xi_0 \xi \tilde{\xi}+
\mathbf{g_6} \xi_0 \tilde{\xi}^2~~~
\label{wd}
\eea
where the mass parameter $\mathbf{M}$ and the coupling constants $\mathbf{g_i}$ are expanded as superfields
\beq
\mathbf{M} = M + \mathbb{a_m} M m_{SUSY} \theta^2
\;\;\; \mbox{and} \;\;\;
\mathbf{g_i} = g_i + \mathbb{g_i} m_{SUSY} \theta^2 \;\;\; (i=0,...,6)
\eeq
and $\mathbb{a_m}$, $g_i$ and $\mathbb{g_i}$ are complex numbers with absolute value of order one. $(\ldots)$ denotes in
eq. (\ref{wd}) the contraction to an $A_4$ invariant. From eq. (\ref{wd}) one can
derive the superpotential given in \cite{af2} at the LO. At the NLO in the expansion parameter $\epsilon$ all non-renormalizable terms
which are suppressed by one power of the cutoff scale $\La_f$ and respect all symmetries of the model are included
(called $\Delta w_d$ in \cite{af2})
\be
w_d^{(1)}=\dd\frac{1}{\Lambda_f}\left(
\sum_{k=3}^{13} \mathbf{t_k} I_k^T+
\sum_{k=1}^{12} \mathbf{s_k} I_k^S+
\sum_{k=1}^{3} \mathbf{x_k} I_k^X
\right)
\ee
where $\{I_k^T,I_k^S,I_k^X\}$ represent a basis of independent quartic invariants
\be
\begin{array}{lp{1in}l}
I_3^T=(\varphi_0^T\varphi_T) (\varphi_T\varphi_T)&&
I_9^T=\left(\varphi_0^T(\varphi_S\varphi_S)_S\right) \xi\\
I_4^T=(\varphi_0^T\varphi_T)' (\varphi_T\varphi_T)''&&
I_{10}^T=\left(\varphi_0^T(\varphi_S\varphi_S)_S\right) \tilde{\xi}\\
I_5^T=(\varphi_0^T\varphi_T)'' (\varphi_T\varphi_T)'&&
I_{11}^T=(\varphi_0^T\varphi_S) \xi^2\\
I_6^T=(\varphi_0^T\varphi_S) (\varphi_S\varphi_S)&&
I_{12}^T=(\varphi_0^T\varphi_S) \xi \tilde{\xi}\\
I_7^T=(\varphi_0^T\varphi_S)' (\varphi_S\varphi_S)''&&
I_{13}^T=(\varphi_0^T\varphi_S) {\tilde{\xi}}^2\\
I_8^T=(\varphi_0^T\varphi_S)'' (\varphi_S\varphi_S)'
\end{array}
\ee
\be
\begin{array}{lp{1in}l}
I_1^S=\left((\varphi_0^S\varphi_T)_S(\varphi_S\varphi_S)_S\right)&&
I_7^S=\left(\varphi_0^S(\varphi_T\varphi_S)_S\right) \tilde{\xi}\\
I_2^S=\left((\varphi_0^S\varphi_T)_A(\varphi_S\varphi_S)_S\right)&&
I_8^S=\left(\varphi_0^S(\varphi_T\varphi_S)_A\right) \xi\\
I_3^S=(\varphi_0^S\varphi_T) (\varphi_S\varphi_S)&&
I_9^S=\left(\varphi_0^S(\varphi_T\varphi_S)_A\right) \tilde{\xi}\\
I_4^S=(\varphi_0^S\varphi_T)' (\varphi_S\varphi_S)''&&
I_{10}^S=(\varphi_0^S\varphi_T) \xi^2\\
I_5^S=(\varphi_0^S\varphi_T)'' (\varphi_S\varphi_S)'&&
I_{11}^S=(\varphi_0^S\varphi_T) \xi \tilde{\xi}\\
I_6^S=\left(\varphi_0^S(\varphi_T\varphi_S)_S\right) \xi&&
I_{12}^S=(\varphi_0^S\varphi_T) {\tilde{\xi}}^2
\end{array}
\ee
\be
\begin{array}{lp{1in}l}
I_1^X=\xi_0 \left(\varphi_T(\varphi_S\varphi_S)_S\right)&&
I_3^X=\xi_0 (\varphi_S\varphi_T) \tilde{\xi}\\
I_2^X=\xi_0 (\varphi_S\varphi_T) \xi
\end{array}
\ee
and the parameters $\mathbf{t_k}$, $\mathbf{s_k}$ and $\mathbf{x_k}$ are expanded in terms of superfields with constant $\theta^2$ component
\beq
\mathbf{t_k}= t_k + \mathbb{t_k} m_{SUSY} \theta^2  \; , \;\;  \mathbf{s_k}= s_k + \mathbb{s_k} m_{SUSY} \theta^2  \;\;\; \mbox{and} \;\;\;
\mathbf{x_k}= x_k + \mathbb{x_k}  m_{SUSY} \theta^2 \; .
\eeq
By $(\ldots)'$ and $(\ldots)''$ the contraction to a non-trivial $A_4$ singlet $1'$ and $1''$ is denoted, respectively.
For the product of two triplets $(\ldots)_{A,S}$ denotes the (anti-) symmetric triplet in this product. 
We agree on the result found in \cite{af2} (up to misprints).

With the information given above we find for the potential $V$ of flavons and driving fields that it is of
the following form
\beq
V= V_{SUSY} + V_{soft}
\eeq
with
\beq
V_{SUSY}= \sum \limits _i \left| \frac{\partial w}{\partial \varphi_i} \right|^2
\eeq
with $\varphi_i$ being in the list $\left\{ \varphi_T, \varphi_S, \xi, \tilde\xi, \varphi^T_0, \varphi^S_0, \xi_0
\right\}$ of all flavons and driving fields.  The relevant part of $V_{soft}$ is given by
\bea
V_{soft}&=&\mathbb{a_m} M m_{SUSY} (\varphi_0^T \varphi_T)+ m_{SUSY} \left[ 
\phantom{\frac{\tilde{g}_4}{2 g_0 \tilde{g}_3}} \!\!\!\!\!\!\!\!\!\!\!\!\!\!\!
\mathbb{g_0} (\varphi_0^T \varphi_T\varphi_T) \right. \nn\\
&&+\left. \mathbb{g}_1 (\varphi_0^S \varphi_S\varphi_S)+
\mathbb{g}_2 \tilde{\xi} (\varphi_0^S \varphi_S)+
\mathbb{g}_3 \xi_0 (\varphi_S\varphi_S)+
\mathbb{g}_4 \xi_0 \xi^2+
\mathbb{g}_5 \xi_0 \xi \tilde{\xi}+
\mathbb{g}_6 \xi_0 \tilde{\xi}^2 \right.\nn\\
&&+\left.  \frac{1}{\Lambda_f}\left(
\sum_{k=3}^{13} \mathbb{t_k} I_k^T+
\sum_{k=1}^{12} \mathbb{s_k} I_k^S+
\sum_{k=1}^{3} \mathbb{x_k} I_k^X
\right) \right]+{\rm h.c.}
\eea
In the course of calculation it is convenient to make the following
re-definition of the couplings $g_3$, $g_4$ and $\mathbb{g_3}$, $\mathbb{g_4}$ in the potential \cite{af2}
\be
\label{redefg3g4}
g_3\equiv 3 \, \tilde{g}_3^2 \; , \;\;\;\; g_4\equiv -\tilde{g}_4^2 \; , \;\;\;
\mathbb{g_3}\equiv 3\, \mathbb{\tilde{g}_3}^2 \;\;\; \mbox{and} \;\;\; \mathbb{g_4}\equiv -\mathbb{\tilde{g}_4}^2 \; .
\ee
We calculated all contributions to the flavon VEVs and VEVs of the driving
fields up to the NLO, that is $\mathcal{O}(\epsilon^2)$ for the flavons and $\mathcal{O}(\epsilon)$ for the driving fields.
Concerning the expansion in $m_{SUSY}$, all flavon VEVs are given at order $\mathcal{O}(m_{SUSY}^0)$
and all VEVs of driving fields are at the LO $\mathcal{O}(m_{SUSY})$.
For the VEVs of the flavons we confirm the result given in \cite{af2}, which we list in the Appendix for completeness.
This result can be cast into the form as given in eq. (\ref{flavonVEVs}) by properly defining the order one parameters.
Apart from the specific structure which is revealed in the LO result, the NLO results only have as special property that the shifted
vacua of $\langle\varphi_{T2}\rangle$ and $\langle\varphi_{T3}\rangle$ coincide, as already noted in \cite{af2}. As we can see from the explicit expression
given in the Appendix, the VEV of one combination of the flavons $\varphi_S$ and $\xi$ is undetermined, indicating the existence of a flat direction in the potential.
Furthermore, note that the vacua of $\varphi_{T2}$, $\varphi_{T3}$ and $\tilde\xi$ only arise at
the NLO level.

Similarly, we find for the vacua of the driving fields
\begin{eqnarray}
\frac{\langle\varphi^T_0\rangle}{m_{SUSY}}&=&c^0_c(1,0,0)+(c^{\prime 0} \epsilon,c^0 \epsilon,c^0 \epsilon) \; , \;\;
\frac{\langle\varphi^S_0\rangle}{m_{SUSY}}= c^0_b(1,1,1)+(c^0_1 \epsilon, c^0_2 \epsilon, c^0_3 \epsilon) \; , \;\; \nonumber \\
\frac{\langle\xi_0\rangle}{m_{SUSY}}&=&c^0_a + c^0_4 \epsilon~~~~~.
\label{drivingVEVs}
\end{eqnarray}
The explicit expressions of the coefficients $c^0$, $c'^0$, $c^0_{a,b,c}$ and $c^0_i$ can be deduced from the explicit expression of the VEVs.
For instance, the VEV of $\varphi_{01}^T$ is given by
\bea
\label{phiT01}
\langle\varphi^T_{01}\rangle &=& m_{SUSY} \, \left[ Y \,
+ \, \left[ \left( \frac{\tilde{g}_4}{2 g_0 \tilde{g}_3}
\left\{ \left( 3 t_{11} + \frac{\tilde{g}_4^2}{\tilde{g}_3^2} (t_6 +t_7 + t_8)\right) Y
+ \frac{1}{2 g_0} \left( 3 (\frac{\mathbb{g_0}}{g_0} t_{11}-\mathbb{t_{11}}) \right.\right.\right.\right.\right. \nn\\
&& \left.\left.\left.\left.\left.+ \frac{\tilde{g}_4^2}{\tilde{g}_3^2} (
\frac{\mathbb{g_0}}{g_0} (t_6 +t_7 + t_8)-(\mathbb{t_{6}}+\mathbb{t_{7}}+\mathbb{t_{8}}))
\right) \right\}  + \frac{3}{8 g_0^2 \tilde{g}_3^3 \tilde{g}_4}\left( t_{11} (3 \tilde{g}_3^2 -2 \tilde{g}_4^2) \right.\right.\right.\right. \nn \\
&& \left.\left.\left.\left. + 3 \tilde{g}_4^2 (t_6+t_7+t_8)\right) Z 
\phantom{\frac{\tilde{g}_4}{2 g_0 \tilde{g}_3}} \!\!\!\!\!\!\!\!\!\!\!\!\!\!\!
\right) \left( \frac{u^3}{v_T^2 \La_f} \right)
 - \frac{9}{2 g_0} \left( t_3 Y + \frac{1}{2 g_0} (\mathbb{t_3}-\frac{\mathbb{g_0}}{g_0} t_3) \right) \left(\frac{v_T}{\La_f} \right)
\right.\right. \nn \\
&& \left.\left. - \frac{g_5}{8 g_0 g_2 \tilde{g}_3 \tilde{g}_4^3} \left( 3 s_{10} + \frac{\tilde{g}_4^2}{\tilde{g}_3^2}
\left( s_3+s_4+s_5 - \frac{g_2}{g_5} x_2 \right) \right) \left( \frac{u^2}{v_T \La_f} \right) Z
\right] \right]
\eea
where
\be
\label{defYZ}
Y = \frac{3}{2 g_0^2} \left( \mathbb{a_m} g_0 - \mathbb{g_0} \right) \;\;\; \mbox{and} \;\;\;
Z =  \frac{\mathbb{\tilde{g}_3}^2 \tilde{g}_4^2 -
\tilde{g}_3^2 \mathbb{\tilde{g}_4}^2}{3 \tilde{g}_3^2 + \tilde{g}_4^2}~~~~~.
\ee
The full expressions of the VEVs of all driving fields are listed in the Appendix.
Given this result one can eventually calculate the vacuum of the auxiliary components of the flavon supermultiplets and 
can find the following structure
\begin{eqnarray}
\frac{1}{\La_f} \left\langle \frac{\partial w}{\partial \varphi_T} \right\rangle &=&
\zeta_T \, m_{SUSY} \, \left\{ (\epsilon,0,0)+(c^{\prime T}_{F} \epsilon^2,c^{T}_{F} \epsilon^2,c^{T}_{F} \epsilon^2) \right\} \nonumber\\
\frac{1}{\La_f} \left\langle \frac{\partial w}{\partial \varphi_S} \right\rangle &=&
\zeta_S \, m_{SUSY} \, \left\{ (\epsilon,\epsilon,\epsilon) + (c^S_{F,1} \epsilon^2, c^S_{F,2} \epsilon^2, c^S_{F,3} \epsilon^2) \right\} \nonumber\\
\frac{1}{\La_f} \left\langle \frac{\partial w}{\partial \xi} \right\rangle &=&
\zeta_\xi \, m_{SUSY} \,  \left( \epsilon + c^\xi \epsilon^2 \right) \nonumber\\
\frac{1}{\La_f} \left\langle \frac{\partial w}{\partial \tilde\xi} \right\rangle &=&
\zeta_{\tilde\xi} \, m_{SUSY} \,  \epsilon^2
\label{FtermVEVs}
\end{eqnarray}
where the LO is proportional to $m_{SUSY} \times \epsilon \, \La_f$. The coefficients $\zeta_{T,S,\xi,\tilde{\xi}}$, $c_F^T$, $c_F^{'T}$, $c_{F,i}^S$ and $c^\xi$
can be computed from eqs. (\ref{flavonVEVs}) and (\ref{drivingVEVs}) and the content of the Appendix.

We see that flavon VEVs and the VEVs of the corresponding auxiliary components have a similar structure. Indeed, at the LO in the $\epsilon$ expansion, 
each flavon VEV has exactly the same orientation
in flavon space as the VEV of the corresponding auxiliary component. In the specific model, we consider, only the flavons $\varphi_T$ contribute
to the charged lepton mass matrix at the LO as well as the NLO, for details see \cite{af2}. Thus, the condition given in eq. (\ref{align}) applies 
and the RL slepton mass terms of eq. (\ref{v}) are diagonal in the basis in which the charged lepton mass matrix is diagonal.
At the NLO, however, this does not hold anymore. At the NLO and for the $A_4$ triplets the condition in eq.(\ref{align}) would require a special relation between the two sets of coefficients
($c_F^T$, $c_F^{'T}$, $c_{F,i}^S$) and ($c$, $c'$, $c_i/c_b$). Such a relation is not natural in our model, since the coefficients
($c$, $c'$, $c_i/c_b$) only depend on the superpotential parameters that remain in the SUSY limit, whereas
($c_F^T$, $c_F^{'T}$, $c_{F,i}^S$) depend on the full set of parameters, including those that describe the soft SUSY breaking terms.
In the case studied here realizing such a relation would be even sufficient in order to eliminate all relevant sources of
flavour violation, associated to this type of contribution, because the subleading contributions in the Yukawa couplings do not induce
further sources of flavour violation.
We found an interesting case where the condition in eq. (\ref{align}) is trivially realized, i.e. with $\alpha=0$.
This is the limit of universal soft SUSY breaking parameters, which is defined as
\bea
&& \mathbb{a_m} = \beta \; , \;\; \mathbb{g_i} = \beta \, g_i \; , \nn \\
&& \mathbb{t_k} = \beta \, t_k \; , \;\; \mathbb{s_k} = \beta \, s_k
\;\;\; \mbox{and} \;\;\; \mathbb{x_k} = \beta \, x_k \; .
\eea
The two re-defined parameters, see eq. (\ref{redefg3g4}), are in the universal limit
\be
\mathbb{\tilde{g}_3} = \sqrt{\beta} \, \tilde{g}_3 \;\;\; \mbox{and} \;\;\; \mathbb{\tilde{g}_4} = \sqrt{\beta} \, \tilde{g}_4~~~~~.
\ee
$\beta$ is a complex number with absolute value of order one. We see from eq.(\ref{defYZ}) that the parameters $Y$ and $Z$
vanish in this limit, as well as all other terms on the right-hand side of eq.(\ref{phiT01}). Thus, $\langle\varphi^T_{01}\rangle$
becomes zero. From the expressions of the VEVs in the Appendix we can see that also all other VEVs of the driving fields are zero 
and as a consequence also the VEVs of the auxiliary components of the flavons vanish.
The case of universal soft SUSY breaking parameters is simple and reduces the number of parameters but, without further specification of the mechanism of SUSY breaking, it is not a natural result in our model.

We notice that non-vanishing VEVs for the driving fields contribute to the $\mu$ term of
the two Higgs doublets $H_u$ and $H_d$. The lowest order operator of this type
in the superpotential $w$, allowed by all symmetries of the model, is
\begin{equation}
(\varphi^T_0 \varphi_T) H_u H_d/\La_f \, .
\end{equation}
It generates a contribution to the $\mu$ term of the order of $m_{SUSY} \times \epsilon$.
The size of such a term is expected to be $\lesssim 50$ GeV for $m_{SUSY} \sim \mathcal{O}(1 \, \rm TeV)$.

\section{Relation to supergravity}
In this section we briefly discuss the constraints that can arise by embedding our setup into the supergravity formalism.
In supergravity SUSY is realized as local symmetry and its breaking is always spontaneous. The consequences for the physics at low energies depend on the specific mechanism
of SUSY breaking. To make contact with our previous discussion, where soft breaking terms are already present at the scale relevant to flavour dynamics, we assume that SUSY is broken
at a large scale in a hidden sector and its breaking is transmitted to the observable one via gravitational effects 
\footnote{In gauge mediation SUSY breaking occurs at a much lower scale where flavour dynamics is already frozen and soft terms are flavour blind, 
up to renormalization group effects.}. 

The hidden sector of the theory contains a gauge singlet chiral supermultiplet $h$ (there can be more, but this does not change our discussion) and the observable sector describes
both, flavons, $\varphi_i$, and matter, $y_a=(l,e^c,H_d,...)$, supermultiplets that we collectively denote by $z_I$. The superfield $h$ can develop a VEV of the order of the Planck scale $M_{Pl}$,
whereas the supermultiplets $z_I$ develop VEVs much smaller than $M_{Pl}$, or no VEV at all. The interactions between hidden and observable sectors take place via the dimensionless
combinations $h/M_{Pl}$. We are interested in the flat limit of the supergravity formalism, in which $M_{Pl}$ is taken to infinity and the gravitino mass $m_{3/2}$
is kept fixed. For a compact explicit discussion we restrict ourselves to the case of canonical K\"ahler potential
\be
K=\vert h\vert^2+\vert z_I\vert^2
\ee
and we parametrize the superpotential $\hat{w}$ as
\be
\hat{w}=\hat{w}_h(h)+\hat{w}_d(h,\varphi)+\hat{w}_m(h,\varphi,y)~~~,
\ee
where $\hat{w}_m$ is a polynomial of third degree in the matter fields $y_a$, as the first term on the right-hand side of eq. (\ref{superp}).
Neglecting gauge interactions, the scalar potential of the theory is given by
\be
V=e^{K/M_{Pl}^2}\left[\vert \hat{F}_h\vert^2+\vert \hat{F}_{z_I}\vert^2-3\frac{\vert \hat{w}\vert^2}{M_{Pl}^2}\right]~~~,
\label{Vsugra}
\ee
where
\be
\hat{F}_h=\frac{\partial \hat{w}}{\partial h}+\overline{h} \frac{\hat{w}}{M_{Pl}^2}~~~,~~~~~~~~~~~~~~~~~~
\hat{F}_{z_I}=\frac{\partial \hat{w}}{\partial z_I}+\overline{z_I} \frac{\hat{w}}{M_{Pl}^2}~~~.
\label{fterms}
\ee
In supergravity SUSY is broken by the non-vanishing VEV of some auxiliary field, $F_h$, $F_{z_I}$ \footnote{We have conventionally chosen to call the flavons $\varphi$ observable fields,
but this does not exclude the possibility that their $F$-terms develop sizable VEVs. This is precisely the point that we would like to discuss in this section. Since $\varphi$ are gauge
singlets that couple to the matter fields $y$ (except for right-handed neutrinos) only through non-renormalizable interaction terms and develop VEVs not much smaller than the cutoff scale of the theory, we could have
included them into the hidden sector as well.}. Assuming a vanishing cosmological constant, when SUSY is broken the gravitino acquires a mass
\be
m_{3/2}=\frac {|\langle w\rangle|}{M_{Pl}^2}
\ee
where we have defined a rescaled superpotential 
\be
w=\langle e^{K/2M_{Pl}^2}\rangle \hat{w}~~~.
\label{rescw}
\ee
We will also make use of the rescaled quantities
\be
F_h=\langle e^{K/2M_{Pl}^2}\rangle \hat{F}_h~~~,~~~~~~~F_{\varphi_i}=\langle e^{K/2M_{Pl}^2}\rangle \hat{F}_{\varphi_i}~~~.
\label{rescF}
\ee
Assuming an appropriate asymptotic behaviour of $\hat{w}$, it is possible to take the flat limit
of $V$ in eq. (\ref{Vsugra}) and derive the soft SUSY breaking terms for the matter fields $y$ \cite{Soni}. These include 
\begin{itemize}
\item
Universal soft scalar masses
\be
m_{3/2}^2 \vert y_a\vert^2~~~.
\ee
\item
Additional bilinear and trilinear soft terms of three different types
\be
m_{3/2}~ w_m~~~,~~~~~~~m_{3/2}~ y_a \frac{\partial w_m}{\partial y_a}~~~,~~~~~~~m_{3/2}~ M_{Pl} \frac{\partial w_m}{\partial h}~~~.
\ee
\item
A contribution from the $F$-terms of the flavon supermultiplets
\be
\langle \overline{F_{\varphi_i}} \rangle\frac{\partial w_m}{\partial \varphi_i}=\left\langle \overline{\frac{\partial w_d}{\partial \varphi_i}}\right\rangle\frac{\partial w_m}{\partial \varphi_i}+m_{3/2}~ \langle\varphi_i\rangle \frac{\partial w_m}{\partial \varphi_i}~~~.
\label{third}
\ee
\end{itemize}
In this paper we have analyzed the first contribution on the right-hand side of the previous equation, in the context of global SUSY.
In supergravity it is more natural to discuss the sum of the two contributions as a whole, since they both arise from the VEV of the $F$-terms
of the flavons, $F_{\varphi_i}$. Moreover, it has been observed that the two contributions can approximately cancel against each other
in a class of supergravity models and then the VEVs of $F_{\varphi_i}$ scale as $m_{3/2}^p$ $(p\ge 2)$ instead of being proportional to $m_{3/2}$ 
\cite{dynamicsupp}.

Setting to zero from the beginning the matter fields $y_a$, the minima of $V$ with vanishing cosmological constant for the remaining fields should obey $\langle V\rangle=0$, $\langle\partial V/\partial h\rangle=0$ and $\langle\partial V/\partial \varphi_i\rangle=0$. These
are equivalent to
\be
\vert F_h\vert^2+\vert F_{\varphi_j}\vert^2-3 m_{3/2}^2 M_{Pl}^2=0~~~,
\label{V0}
\ee
\be
\left(\frac{\partial^2 w}{\partial h^2}+\frac{\overline{h}}{M_{Pl}^2}\frac{\partial w}{\partial h}\right)\overline{F_h}
+\left(F_h -3\frac{\partial w}{\partial h}\right)m_{3/2}+
\left(\frac{\partial^2 w}{\partial h\partial \varphi_i}+\frac{\overline{\varphi_i}}{M_{Pl}^2}\frac{\partial w}{\partial h}\right)\overline{F_{\varphi_i}}=0~~~,
\label{dVh}
\ee
\be
\left(\frac{\partial^2 w}{\partial \varphi_i \partial h}+\frac{\overline{h}}{M_{Pl}^2}\frac{\partial w}{\partial \varphi_i}\right)\overline{F_h}
+\left(F_{\varphi_i} -3\frac{\partial w}{\partial \varphi_i}\right)m_{3/2}+
\left(\frac{\partial^2 w}{\partial \varphi_i\partial \varphi_j}+\frac{\overline{\varphi_j}}{M_{Pl}^2}\frac{\partial w}{\partial \varphi_i}\right)\overline{F_{\varphi_j}}=0~~~,
\label{dVphi}
\ee
\vskip 0.2 cm
\noindent
where we have used the rescaled functions of eqs.  (\ref{rescw}) and (\ref{rescF}). 
Our aim is to analyze the behaviour of the previous equations in the limit of small $m_{3/2}$, by performing a series expansion in powers of $m_{3/2}$. From eq. (\ref{V0}) we see that the $F$-terms should 
vanish when we take $m_{3/2}$ to zero and some of them should scale as $m_{3/2}$ at the minimum: we assume that $F_h\propto m_{3/2}$.
We also assume that at the minimum of $V$, $h$ as well as some of the fields $\varphi_i$ tend to some non-vanishing constant in the limit of vanishing $m_{3/2}$. 
In realistic models $h$ is of order $M_{Pl}$ at the minimum, and the flavon fields have a VEV much larger than $m_{3/2}$, though smaller than $M_{Pl}$.
We now impose that also the VEV of $F_{\varphi_i}$ scales as $m_{3/2}$ in the limit of small gravitino mass
\be
F_{\varphi_i}\propto m_{3/2}~~~.
\ee
We analyze the conditions under which such behaviour is consistent with eqs. (\ref{dVh},\ref{dVphi}). From eq. (\ref{fterms}) we see that we also 
have $\partial w/\partial h\propto m_{3/2}$ and $\partial w/\partial \varphi_i\propto m_{3/2}$ at maximum. Then it is easy to see that all the terms of eqs. (\ref{dVh},\ref{dVphi}) not containing second derivatives of $w$
scale at least as $m_{3/2}^2$. The equations can be satisfied if either of the following two cases occurs.
\begin{itemize}
\item[\bf I)]
In the first case we have
\be
\frac{\partial^2 w}{\partial \varphi_i \partial h}\propto m_{3/2}^p~~~~~~~~~~p\ge 1~~~,
\ee
(including the case $\partial^2 w/\partial \varphi_i\partial h = 0$). 
Then eq. (\ref{dVh}) can be satisfied, for instance, by $\partial^2 w/\partial h^2=0$ as in the Polonyi model and eq. (\ref{dVphi}) requires
\be
\frac{\partial^2 w}{\partial \varphi_i\partial \varphi_j}\propto m_{3/2}^p~~~~~~~~~~p\ge 1~~~.
\ee
At first sight it may seem surprising or unnatural that at the minimum $\partial^2 w/\partial \varphi_i\partial \varphi_j$ vanish in the limit of zero $m_{3/2}$, since
this matrix controls the flavon masses and its dynamics is expected to be related to a much higher scale. This tuning is however no worse than that
occurring in the Polonyi model, where $F_h$ is of order $m_{3/2} M_{Pl}$ (instead of $M_{Pl}^2$) despite the VEV for $h$ of order $M_{Pl}$.
In the Polonyi model this is obtained by tuning by hand the overall scale of the superpotential, 
$\hat{w}_h=m M_{Pl} (h+b M_{Pl})$ with $m$ of order $m_{3/2}$.  
Therefore a minimum with $F_{\varphi_i}$ proportional to $m_{3/2}$ can occur, if the superpotential is of the type
\be
\hat{w}=m \left(M_{Pl} h+b M_{Pl}^2+a_{ij} \varphi_i\varphi_j+\frac{g_{ijk}}{M_{Pl}}\varphi_i\varphi_j\varphi_k\right)
\ee
and we have checked in an explicit example that $\langle F_{\varphi_i}\rangle$ can be of the order $m_{3/2} \times \langle \varphi_i\rangle$.

Notice that this case also includes the possibility where $\partial^2 w/\partial \varphi_i\partial h$ and $\partial^2 w/\partial \varphi_i\partial \varphi_j$ vanish at the minimum, which can occur if the superpotential does not depend on the flavon fields $\varphi_i$, or depends on them in combination with fields having vanishing VEVs. Eq. (\ref{dVh}) can then be satisfied by a superpotential linear in the hidden sector field $h$, as in the Polonyi model.

\item[\bf II)]
In the second case
\be
\frac{\partial^2 w}{\partial \varphi_i \partial h}\ne 0~~~~~~~{\rm for~ vanishing}~ m_{3/2} \; .
\ee
Barring cancellations, in this case also the remaining second derivatives, $\partial^2 w/\partial h^2$ and $\partial^2 w/\partial \varphi_i\partial \varphi_j$, should be non-vanishing when $m_{3/2}$ tends to zero and 
the terms containing second derivatives should cancel against each other in the equations. When this happens, the couplings between the supermultiplets $h$ and $\varphi_i$ are large
and it would be more appropriate to include $\varphi_i$ into the hidden sector. We do not know examples of this type among the most common superpotentials considered in supergravity,
but we think that it is not possible to discard a priori this possibility.
\end{itemize}

We also observe that, if there is no coupling between the hidden sector and the flavon fields, $\partial^2 w/\partial \varphi_i \partial h=0$, and if $\partial^2 w/\partial \varphi_i\partial \varphi_j$
is non-vanishing in the limit of zero $m_{3/2}$, then eq. (\ref{dVphi}) can only be solved if $F_{\varphi_i}\propto m_{3/2}^p$, with $p\ge 2$, 
i.e. the two contributions to $F_{\varphi_i}$ have to cancel up to terms of order $m_{3/2}^p$
\footnote{ Notice that this case is different from the one included under condition {\bf I)} above, where the vanishing of $\partial^2 w/\partial \varphi_i\partial h$ is considered as possible solution, but 
$\partial^2 w/\partial \varphi_i\partial \varphi_j$ is assumed to fulfill
$\partial^2 w/\partial \varphi_i\partial \varphi_j\propto m_{3/2}^p~~(p\ge 1)$ .}.
In this case, the contribution shown in eq. (\ref{third}),
which is the subject of the present work, is harmless, as has been noticed in ref. \cite{dynamicsupp}. It is quite interesting that in this class of models the suppression of this contribution occurs dynamically
through the minimization of the scalar potential of the underlying supergravity theory. 

The framework considered in this section is not the most general one. We could also allow for a non-canonical K\"ahler potential, a possibility that leads to soft SUSY breaking terms
and minimum conditions more general than those analyzed here. 
In the setup of global SUSY which we have analyzed in the previous sections, we wished to contemplate the most general possibility, without making any assumption
about the origin and the specific pattern of the SUSY breaking terms. Even if cancellations in the VEVs of the $F$-terms for the flavon fields can occur
and do occur in specific cases, as we have seen these cancellations are not model-independent features of the underlying supergravity theory and the general parametrization of our global framework
is more appropriate to cover the most general possibility.

\section{Summary and Conclusions}
In this note we have studied the effect of generic soft SUSY breaking terms
on the vacuum alignment in a globally SUSY model invariant under the flavour symmetry
$A_4\times Z_3\times \Uu_{FN}\times \Uu_R$.
In such a model, lepton masses and mixing angles directly depend
on how the flavour symmetry is broken by the flavon fields.
In the SUSY limit the minimization of the scalar potential of the theory
leads to a special pattern of flavon VEVs $\langle\varphi\rangle$, that reproduces the nearly tri-bimaximal mixing observed in neutrino oscillations.
The question addressed here is how this vacuum structure is modified when generic soft SUSY breaking terms
are added to a globally SUSY theory.
At first sight the impact of such terms would seem negligible, due to the large
separation between the flavour symmetry breaking scale, $\langle\varphi\rangle\approx 10^{14}$ GeV
and $m_{SUSY}\approx 1$ TeV. Indeed the corrections to the VEVs of the flavon fields
induced by the soft SUSY breaking terms are of order $m_{SUSY}$ and thus completely
irrelevant as far as lepton masses and mixing angles are concerned.

Even if lepton masses and mixings are unaffected by the soft SUSY breaking terms,
there are important corrections to the VEVs of the auxiliary components
of the flavon supermultiplets. These are zero in the SUSY limit, and
become non-vanishing when soft SUSY breaking terms are included. By an explicit computation we find
that $\langle\partial w/\partial \varphi\rangle$ are of order $m_{SUSY} \times \langle\varphi\rangle$.
These VEVs give rise to a contribution to the RL slepton masses of order
$m_i m_{SUSY}$, $m_i$, $i=e,\mu,\tau$, denoting the lepton masses.
The important feature of this contribution
is its orientation in flavour space, which is completely determined at the LO by the relative orientation
of $\langle\ov{\partial w/\partial \varphi}\rangle$  with respect to $\langle\varphi\rangle$.
A misalignment is a source of lepton flavour violation, since it gives rise to non-diagonal terms
in the RL slepton mass matrix in the basis in which the charged lepton mass matrix is diagonal.
In our model we can compute both $\langle\partial w/\partial \varphi\rangle$ and $\langle\varphi\rangle$
in a systematic expansion in the parameter $\epsilon$, the scale at which the flavour symmetry
is broken measured in units of the cutoff scale $\Lambda_f$.
At the LO in this expansion, for each irreducible multiplet $\varphi$ of the flavour symmetry
we find $\langle\ov{\partial w/\partial \varphi}\rangle\propto \langle\varphi\rangle$. At the NLO however,
by including terms of ${\cal O}(\epsilon^2 \Lambda_f)$ in $\langle\varphi\rangle$
and terms of ${\cal O}(\epsilon^2 m_{SUSY} \Lambda_f)$ in $\langle\partial w/\partial \varphi\rangle$,
such a proportionality does not hold any longer. The VEVs of the flavons and of their auxiliary components are misaligned
and there are non-diagonal contributions to the RL slepton mass matrix in the flavour basis.
\footnote{Subleading terms present in the Yukawa couplings might be an additional source of flavour violation associated to 
this type of contribution to the RL slepton masses.}

By inspecting the explicit expressions of $\langle\partial w/\partial \varphi\rangle$,
we have found a special case in which they vanish. This occurs when
the soft SUSY breaking terms of the flavons are universal, that is they have the same form,
up to an overall proportionality constant, as the superpotential terms.
In our model however this special case is not a generic result and thus has to be considered as fine-tuning, as
long as the mechanism of SUSY breaking is not specified.
As has been discussed in \cite{RV} in the context of supergravity, the VEVs of the $F$-terms 
of the flavons are also generically expected to be of order $m_{3/2} \times \langle \varphi_i \rangle \sim 
m_{SUSY} \times \langle \varphi_i \rangle$ and thus give a relevant contribution
to the sfermion soft masses of RL type. However, these terms can be suppressed \cite{dynamicsupp} 
so that they are $\ll m_{3/2} \times \langle \varphi_i \rangle$ 
through a dynamical mechanism in which the generic supergravity contribution is canceled against the globally SUSY one.
 In this work we have recovered in a model-independent way the conditions under which such a dynamical suppression occurs 
by performing a series expansion of the relevant minimum conditions in powers of $m_{3/2}$; for an explicit
example see \cite{dynamicsupp}.

One relevant consequence of our result concerns processes in which lepton flavour is
violated, especially $\mu\to e \gamma$, whose branching ratio is severely constrained.
Actually it is possible to show that the amplitude for $l_i\to l_j \gamma$ is dominated regarding the expansion in the symmetry breaking parameter $\epsilon$
by the above mechanism, through a one-loop diagram with the insertion of the element $ij$ of
the RL block of the slepton mass matrix. 
In particular the normalized branching ratios $R_{ij}$ for the lepton flavour violating transitions
$l_i\to l_j \gamma$
\be
R_{ij}=\frac{BR(l_i\to l_j\gamma)}{BR(l_i\to l_j\nu_i{\bar \nu_j})}~~~,
\ee
have the following asymptotic behaviour for small $\epsilon$
\be
R_{ij}=\frac{48\pi^3 \alpha_{em}}{G_F^2 M_{new}^4}\vert w_{ij} ~\epsilon\vert^2
\label{LFV}
\ee
where $\alpha_{em}$ is the fine structure constant, $G_F$ is the Fermi constant,
$w_{ij}$ are dimensionless parameters of order one and $M_{new}=(4\pi/g)m_{SUSY}$ with $g$ being the $SU(2)_L$ gauge coupling constant.
When the contribution to the RL slepton masses from the $F$-term of the flavon multiplets is absent or negligible,
a cancellation takes place in the amplitudes for lepton flavour violating transitions and $R_{ij}$
scale as
\be
R_{ij}= \frac{48\pi^3 \alpha_{em}}{G_F^2 M_{new}^4}\left[\vert w^{(1)}_{ij} \epsilon^2\vert^2+\frac{m_j^2}{m_i^2} \vert w^{(2)}_{ij} \epsilon\vert^2\right]
\label{LFVsusy}
\ee
where $w^{(1,2)}_{ij}$ are dimensionless quantities of order one. 
Given the smallness of the symmetry breaking parameter $\epsilon$, the branching ratios in eq. (\ref{LFVsusy}) are clearly
much more suppressed than those in eq. (\ref{LFV}). This shows the potential relevance of the effect analyzed in this paper.
A detailed calculation of the branching ratios
for radiative charged lepton decays will be presented elsewhere \cite{fhlm09}.

In this note we discussed a model in the framework of global SUSY and we carried out an explicit computation of the effect up to
the NLO in the parameter $\epsilon$ showing, without specifying the SUSY breaking mechanism, the relevance
of the $F$-terms of the flavon fields in model building.
In our specific framework, a contribution to the $\mu$ term is generated as well due to the VEVs of the driving fields which
are of the size of the generic soft SUSY breaking scale $m_{SUSY}$.

\section*{Acknowledgments}

We thank Fabio Zwirner for useful comments.
We recognize that this work has been partly supported by the European Commission under contract MRTN-CT-2006-035505 and partly by
the European Programme "Unification in the LHC Era", contract PITN-GA-2009-237920 (UNILHC).

\newpage
\appendix
\mathversion{bold}
\section{NLO vacua for flavons and driving fields}
\mathversion{normal}
Here we list the explicit expressions of the VEVs of flavon and driving fields at the NLO in the expansion parameter $\epsilon$.
Concerning the expansion in the soft SUSY breaking parameter $m_{SUSY}$, the results are given at the LO,
namely at ${\cal O}(m_{SUSY}^0)$ for the flavons and at  ${\cal O}(m_{SUSY})$ for the driving fields.
For the flavons we find
\bea
\langle\varphi_{T1}\rangle &=& -\frac{3 M}{2 g_0} \, - \, \left[ \frac{3 t_3}{2 g_0} \left( \frac{v_T^2}{\La_f} \right)\,
+ \, \frac{\tilde{g}_4}{2 g_0 \tilde{g}_3} \left( t_{11} + \frac{\tilde{g}_4^2}{3 \tilde{g}_3^2} \left( t_6 + t_7 + t_8\right) \right) \,
\left( \frac{u^3}{v_T \, \La_f} \right) \right]\nn\\
\langle\varphi_{T2}\rangle &=&
\frac{\tilde{g}_4}{4 g_0 \tilde{g}_3} \left[ t_{11} + \frac{\tilde{g}_4^2}{3 \tilde{g}_3^2} \left( t_6 + t_7 + t_8\right) \right] \,
\left( \frac{u^3}{v_T \, \La_f} \right)\nn\\
\langle\varphi_{T3}\rangle &=&
\frac{\tilde{g}_4}{4 g_0 \tilde{g}_3} \left[ t_{11} + \frac{\tilde{g}_4^2}{3 \tilde{g}_3^2} \left( t_6 + t_7 + t_8\right) \right] \,
\left( \frac{u^3}{v_T \, \La_f} \right)\nn\\
\langle\varphi_{S1}\rangle &=& \frac{\tilde{g}_4 \, u}{3 \tilde{g}_3} \, + \, \left[ \frac{1}{\tilde{g}_4}
\left( -\frac{\tilde{g}_3}{g_1} + \frac{g_5}{6 g_2 \tilde{g}_4} \right) s_{10}
- \frac{\tilde{g}_4}{6 g_1 \tilde{g}_3} \left( 2 s_3 - s_4 - s_5\right)\right.\nn\\
&&\left.+ \frac{g_5}{18 g_2 \tilde{g}_3^2} \left( s_3 + s_4 + s_5 \right) - \frac{s_6}{3 g_1} - \frac{x_2}{18 \tilde{g}_3^2}
\right] \left(\frac{u v_T}{\La_f} \right)\nn\\
\langle\varphi_{S2}\rangle &=& \frac{\tilde{g}_4 \, u}{3 \tilde{g}_3} \, + \, \left[ \frac{1}{\tilde{g}_4}
\left( \frac{\tilde{g}_3}{2 g_1} + \frac{g_5}{6 g_2 \tilde{g}_4} \right) s_{10}
- \frac{\tilde{g}_4}{6 g_1 \tilde{g}_3} \left( 2 s_4 - s_3 - s_5\right)\right.\nn\\
&&\left.+ \frac{g_5}{18 g_2 \tilde{g}_3^2} \left( s_3 + s_4 + s_5 \right) + \frac{s_6}{6 g_1} + \frac{s_8}{4 g_1}  - \frac{x_2}{18 \tilde{g}_3^2}
\right] \left(\frac{u v_T}{\La_f} \right)\nn\\
\langle\varphi_{S3}\rangle &=& \frac{\tilde{g}_4 \, u}{3 \tilde{g}_3} \, + \,  \left[ \frac{1}{\tilde{g}_4}
\left( \frac{\tilde{g}_3}{2 g_1} + \frac{g_5}{6 g_2 \tilde{g}_4} \right) s_{10}
- \frac{\tilde{g}_4}{6 g_1 \tilde{g}_3} \left( 2 s_5 - s_3 - s_4\right)\right.\nn\\
&&\left.+ \frac{g_5}{18 g_2 \tilde{g}_3^2} \left( s_3 + s_4 + s_5 \right) + \frac{s_6}{6 g_1} - \frac{s_8}{4 g_1}  - \frac{x_2}{18 \tilde{g}_3^2}
\right] \left(\frac{u v_T}{\La_f} \right)\nn\\
\langle\xi\rangle &=& u\nn\\
\langle\tilde\xi\rangle &=& -\left[ \frac{\tilde{g}_{3} s_{10}}{g_2 \tilde{g}_4}
+ \frac{\tilde{g}_4}{3 g_2 \tilde{g}_3} \left( s_3 + s_4 + s_5\right) \right] \left( \frac{u v_T}{\La_f} \right)
\eea
$v_T$ is defined as
\be
v_T = -\frac{3 M}{2 g_0} \, ,
\ee
equal to the VEV of the first component of the flavon field $\varphi_{T}$ at the LO. Note that the parameter $u$ is undetermined in the VEVs.
\newpage
With the parameters $Y$ and $Z$, as given in eq.(\ref{defYZ}), we get for the VEVs of the driving fields
\bea
\langle\varphi^T_{01}\rangle &=& m_{SUSY} \, \left[ Y \,
+ \, \left[ \left( \frac{\tilde{g}_4}{2 g_0 \tilde{g}_3}
\left\{ \left( 3 t_{11} + \frac{\tilde{g}_4^2}{\tilde{g}_3^2} (t_6 +t_7 + t_8)\right) Y
+ \frac{1}{2 g_0} \left( 3 (\frac{\mathbb{g_0}}{g_0} t_{11}-\mathbb{t_{11}}) \right.\right.\right.\right.\right. \nn\\
&& \left.\left.\left.\left.\left.+ \frac{\tilde{g}_4^2}{\tilde{g}_3^2} (
\frac{\mathbb{g_0}}{g_0} (t_6 +t_7 + t_8)-(\mathbb{t_{6}}+\mathbb{t_{7}}+\mathbb{t_{8}}))
\right) \right\}  + \frac{3}{8 g_0^2 \tilde{g}_3^3 \tilde{g}_4}\left( t_{11} (3 \tilde{g}_3^2 -2 \tilde{g}_4^2) \right.\right.\right.\right. \nn \\
&& \left.\left.\left.\left. + 3 \tilde{g}_4^2 (t_6+t_7+t_8)\right) Z \phantom{\frac{\tilde{g}_4}{2 g_0 \tilde{g}_3}} \!\!\!\!\!\!\!\!\!\!\!\!\!\!\!
\right) \left( \frac{u^3}{v_T^2 \La_f} \right)
 - \frac{9}{2 g_0} \left( t_3 Y + \frac{1}{2 g_0} (\mathbb{t_3}-\frac{\mathbb{g_0}}{g_0} t_3) \right) \left(\frac{v_T}{\La_f} \right)
\right.\right. \nn \\
&& \left.\left. - \frac{g_5}{8 g_0 g_2 \tilde{g}_3 \tilde{g}_4^3} \left( 3 s_{10} + \frac{\tilde{g}_4^2}{\tilde{g}_3^2}
\left( s_3+s_4+s_5 - \frac{g_2}{g_5} x_2 \right) \right) \left( \frac{u^2}{v_T \La_f} \right) Z
\right] \right] \nn \\
\langle\varphi^T_{02}\rangle &=& m_{SUSY}  \left[ \left( \frac{\tilde{g}_4}{16 g_0^2 \tilde{g}_3}
\left( 3 (\frac{\mathbb{g_0}}{g_0} t_{11} -\mathbb{t_{11}}) + \frac{\tilde{g}_4^2}{\tilde{g}_3^2} (\frac{\mathbb{g_0}}{g_0} (t_6 +t_7 + t_8)
-(\mathbb{t_{6}}+\mathbb{t_{7}}+\mathbb{t_{8}})) \right) \right.\right. \nn\\
&& \left.\left. + \frac{3}{32 g_0^2 \tilde{g}_3^3 \tilde{g}_4}\left( t_{11} (3 \tilde{g}_3^2 -2 \tilde{g}_4^2) + 3 \tilde{g}_4^2 (t_6+t_7+t_8)\right) Z
\right) \left( \frac{u^3}{v_T^2 \La_f} \right) \right.\nn \\
&& \left. + \frac{g_5}{16 g_0 g_2 \tilde{g}_3 \tilde{g}_4^3} \left( 3 s_{10} + \frac{\tilde{g}_4^2}{\tilde{g}_3^2} \left( s_3+s_4+s_5 - \frac{g_2}{g_5} x_2
\right) \right) \left( \frac{u^2}{v_T \La_f} \right) Z \right] \nn \\
\langle\varphi^T_{03}\rangle &=& m_{SUSY} \left[ \left( \frac{\tilde{g}_4}{16 g_0^2 \tilde{g}_3}
\left( 3 (\frac{\mathbb{g_0}}{g_0} t_{11} -\mathbb{t_{11}}) + \frac{\tilde{g}_4^2}{\tilde{g}_3^2} (\frac{\mathbb{g_0}}{g_0} (t_6 +t_7 + t_8)
-(\mathbb{t_{6}}+\mathbb{t_{7}}+\mathbb{t_{8}})) \right) \right.\right. \nn\\
&& \left.\left. + \frac{3}{32 g_0^2 \tilde{g}_3^3 \tilde{g}_4}\left( t_{11} (3 \tilde{g}_3^2 -2 \tilde{g}_4^2) + 3 \tilde{g}_4^2 (t_6+t_7+t_8)\right) Z
\right) \left( \frac{u^3}{v_T^2 \La_f} \right) \right.\nn \\
&& \left. + \frac{g_5}{16 g_0 g_2 \tilde{g}_3 \tilde{g}_4^3} \left( 3 s_{10} + \frac{\tilde{g}_4^2}{\tilde{g}_3^2} \left( s_3+s_4+s_5 - \frac{g_2}{g_5} x_2
\right) \right) \left( \frac{u^2}{v_T \La_f} \right) Z \right] \nn \\
\langle\xi_0\rangle &=& m_{SUSY} \, \left[ -\frac{1}{4 \tilde{g}_3^2 \tilde{g}_4^2} Z \, + \, \left[ -\frac{1}{6 \tilde{g}_3 \tilde{g}_4 (3 \tilde{g}_3^2
+ \tilde{g}_4^2)} \left( \phantom{\frac{\tilde{g}_4}{2 g_0 \tilde{g}_3}} \!\!\!\!\!\!\!\!\!\!\!\!\!\!\!
(3 \tilde{g}_3^2 -2 \tilde{g}_4^2) t_{11} + 3 \tilde{g}_4^2 (t_6+t_7+t_8) \right.\right.\right. \nn \\
&& \left.\left.\left.
- \frac{g_0 g_5}{3 g_2} \left( \frac{3 \tilde{g}_3^2}{\tilde{g}_4^2} s_{10} + s_3+s_4+s_5 - \frac{g_2}{g_5} x_2 \right) \left( \frac{v_T}{u} \right)
\right)  \left( \frac{u}{\La_f} \right) Y \right.\right. \nn \\
&& \left.\left.+ \frac{1}{12 g_2^2 \tilde{g}_3 \tilde{g}_4 (3 \tilde{g}_3^2 + \tilde{g}_4^2)} \left(\left( \frac{3 \tilde{g}_3^2}{\tilde{g}_4^2} s_{10} + s_3+s_4+s_5\right) \left( g_2 \mathbb{g_5}  - \mathbb{g_2} g_5 \right)
\left( \frac{v_T}{\La_f} \right) \right.\right.\right. \nn \\
&& \left.\left.\left. + g_2 g_5 \left\{ \left( \mathbb{s_3}+\mathbb{s_4}+\mathbb{s_5}
- \frac{g_2}{g_5} \mathbb{x_2}\right)
- \frac{3 \mathbb{\tilde{g}_3}^2 + \mathbb{\tilde{g}_4}^2}{3 \tilde{g}_3^2 + \tilde{g}_4^2} \left( s_3+s_4+s_5 - \frac{g_2}{g_5} x_2 \right)\right\}
\left( \frac{v_T}{\La_f} \right)\right)\right.\right. \nn \\
&& \left.\left. + \frac{\tilde{g}_3 g_5}{4 g_2 \tilde{g}_4^3 (3 \tilde{g}_3^2 + \tilde{g}_4^2)} \left( \mathbb{s_{10}}
- \frac{\mathbb{\tilde{g}_4}^2}{\tilde{g}_4^2} s_{10} \right) \left( \frac{v_T}{\La_f} \right) + \frac{(3 \tilde{g}_3^2 +2 \tilde{g}_4^2) g_5}{4 g_2 \tilde{g}_3
\tilde{g}_4^5 (3 \tilde{g}_3^2 + \tilde{g}_4^2)} \, s_{10} \, 
\left( \frac{v_T}{\La_f} \right) Z
\right] \right] \nn 
\eea

\bea
\langle\varphi^S_{01}\rangle &=& m_{SUSY} \, \left[ \frac{g_5}{4 g_2 \tilde{g}_3 \tilde{g}_4^3} Z \, + \, \left[ \frac{\tilde{g}_4}{6 g_1 \tilde{g}_3} \left(
-2 t_6 + t_7 + t_8 - \frac{4 \tilde{g}_3}{\tilde{g}_4} t_9 - \frac{6 \tilde{g}_3^2}{\tilde{g}_4^2} t_{11} - \frac{2 g_1 \tilde{g}_3}{g_2 \tilde{g}_4} t_{12}
\right) \left( \frac{u}{\La_f} \right)  Y  \nn \right.\right.\\
&& \left.\left. + \frac{g_5}{2 g_2 (3 \tilde{g}_3^2 + \tilde{g}_4^2)} \left( t_6 + t_7 + t_8 + \frac{(3 \tilde{g}_3^2 -2 \tilde{g}_4^2)}{3 \tilde{g}_4^2} t_{11}
\right) \left( \frac{u}{\La_f} \right)  Y  - \frac{g_0 \tilde{g}_3}{3 g_1^2 \tilde{g}_4} s_6 \left( \frac{v_T}{\La_f} \right) Y \right.\right.\nn \\
&& + \left.\left. \frac{g_0}{18 g_1^2 g_2^2} \left(-2 (2 g_1^2 + 3 g_2^2) s_3 -  (4 g_1^2 - 3 g_2^2) (s_4 +s_5) - \frac{6 \tilde{g}_3^2}{\tilde{g}_4^2} (2 g_1^2+3 g_2^2)
s_{10} \right) \left( \frac{v_T}{\La_f} \right) Y \right.\right. \nn \\
&& + \left.\left. \frac{g_0 g_5^2}{18 g_2^2 \tilde{g}_4^2 (3 \tilde{g}_3^2 + \tilde{g}_4^2)} \left( -s_3 -s_4-s_5 + \frac{g_2}{g_5} x_2
-\frac{3 \tilde{g}_3^2}{\tilde{g}_4^2} s_{10} \right) \left( \frac{v_T}{\La_f} \right) Y
\right.\right. \nn \\
&& \left.\left. + \frac{1}{6 g_1^2 g_2^2} \left(\frac{1}{g_1 g_2} (3 \mathbb{g_1} g_2^3 + 2 g_1^3 \mathbb{g_2}) s_3
 - (2 g_1^2 + 3 g_2^2) \mathbb{s_3} \right) \left( \frac{v_T}{\La_f} \right) \right.\right. \nn \\
&& \left.\left.+ \frac{1}{12 g_1^2 g_2^2} \left(\frac{1}{g_1 g_2} (4 g_1^3 \mathbb{g_2} - 3 \mathbb{g_1} g_2^3) s_4
 - (4 g_1^2 - 3 g_2^2) \mathbb{s_4} \right) \left( \frac{v_T}{\La_f} \right)
\right.\right. \nn \\
&& \left.\left.+ \frac{1}{12 g_1^2 g_2^2} \left(\frac{1}{g_1 g_2} (4 g_1^3 \mathbb{g_2} - 3 \mathbb{g_1} g_2^3) s_5
 - (4 g_1^2 - 3 g_2^2) \mathbb{s_5} \right) \left( \frac{v_T}{\La_f} \right)
\right.\right. \nn \\
&& \left.\left. + \frac{\tilde{g}_3^2}{2 g_1^2 g_2^2 \tilde{g}_4^2} \left(\frac{1}{g_1 g_2} (3 \mathbb{g_1} g_2^3 + 2 g_1^3 \mathbb{g_2}) s_{10}
- (2 g_1^2+3 g_2^2) \mathbb{s_{10}} \right) \left( \frac{v_T}{\La_f} \right)
\right.\right. \nn \\
&& \left.\left. - \frac{\tilde{g}_3^2 g_5}{4 g_2^3 \tilde{g}_4^4 (3 \tilde{g}_3^2 + \tilde{g}_4^2)} \left( g_2 \mathbb{g_5} - \mathbb{g_2} g_5 \right)
s_{10} \left( \frac{v_T}{\La_f} \right) + \frac{\tilde{g}_3^2 g_5^2}{4 g_2^2 \tilde{g}_4^4 (3 \tilde{g}_3^2 + \tilde{g}_4^2)}
\left( \frac{\mathbb{\tilde{g}_4}^2}{\tilde{g}_4^2} s_{10} - \mathbb{s_{10}} \right) \left( \frac{v_T}{\La_f} \right)
\right.\right. \nn \\
&&\left.\left. + \frac{g_5^2}{12 g_2^2 \tilde{g}_4^2 (3 \tilde{g}_3^2 + \tilde{g}_4^2)} \left(\left(\frac{3 \mathbb{\tilde{g}_3}^2 + \mathbb{\tilde{g}_4}^2}{3 \tilde{g}_3^2 + \tilde{g}_4^2}
s_3 - \mathbb{s_3} \right) + \left(\frac{3 \mathbb{\tilde{g}_3}^2
+ \mathbb{\tilde{g}_4}^2}{3 \tilde{g}_3^2 + \tilde{g}_4^2} s_4 - \mathbb{s_4} \right) \right)\left( \frac{v_T}{\La_f} \right)
\right.\right. \nn \\
&& \left.\left.+ \frac{g_5^2}{12 g_2^2 \tilde{g}_4^2 (3 \tilde{g}_3^2 + \tilde{g}_4^2)} \left(\frac{1}{2} \left( \frac{\mathbb{\tilde{g}_4}^2}{\tilde{g}_4^2}
+ \frac{3 \mathbb{\tilde{g}_3}^2 + \mathbb{\tilde{g}_4^2}}{3 \tilde{g}_3^2 + \tilde{g}_4^2} \right) (s_5 - \frac{g_2}{g_5} x_2)  - (\mathbb{s_5} - \frac{g_2}{g_5}
\mathbb{x_2}) \right) \left( \frac{v_T}{\La_f} \right) \right.\right.\nn \\
&& \left.\left. + \frac{g_5}{12 g_2^3 \tilde{g}_4^2 (3 \tilde{g}_3^2 +\tilde{g}_4^2)} \left( \mathbb{g_2} g_5- g_2 \mathbb{g_5}\right) (s_3+s_4+s_5)
\left( \frac{v_T}{\La_f} \right)
+ \frac{\tilde{g}_3}{2 g_1^2 \tilde{g}_4} \left( \frac{\mathbb{g_1}}{g_1} s_6 - \mathbb{s_6} \right) \left( \frac{v_T}{\La_f} \right) \right.\right. \nn \\
&& \left.\left. - \frac{g_5}{6 g_1 g_2 \tilde{g}_3 \tilde{g}_4^3} s_1 \left( \frac{v_T}{\La_f} \right) Z
- \frac{g_5}{4 g_2^2 \tilde{g}_4^4} s_{11} \left( \frac{v_T}{\La_f} \right) Z \right.\right. \nn\\
&& \left.\left. + \frac{1}{6 g_1 \tilde{g}_3^2 \tilde{g}_4^2} x_{1} \left( \frac{v_T}{\La_f} \right) Z
+ \frac{1}{12 g_2 \tilde{g}_3^2 \tilde{g}_4^2} x_{3} \left( \frac{v_T}{\La_f} \right) Z 
\right.\right.\nn \\
&& \left.\left. + \frac{1}{4 g_1^2 \tilde{g}_3 \tilde{g}_4^3} (-(3 \tilde{g}_3^2 +\tilde{g}_4^2) s_6 + g_1 x_2) \left( \frac{v_T}{\La_f} \right) Z
+ \frac{1}{2 g_2^2 \tilde{g}_4^2} \left(s_3 + s_4 + s_5 -s_{10}\right) \left( \frac{v_T}{\La_f} \right) Z \right.\right. \nn \\
&& \left.\left. + \frac{g_5}{8 g_1 g_2 \tilde{g}_3 \tilde{g}_4^3} \left(2 s_3 -s_4 -s_5 + \frac{6 \tilde{g}_3^2}{\tilde{g}_4^2} s_{10}\right)
\left( \frac{v_T}{\La_f}\right) Z \right.\right. \nn \\
&& \left.\left.+ \frac{g_5^2}{24 g_2^2 \tilde{g}_3^2 \tilde{g}_4^4} \left(-s_3-s_4 -s_5 +\frac{g_2}{g_5} x_2\right) \left( \frac{v_T}{\La_f}\right) Z
\right.\right. \nn \\
&& \left.\left.- \frac{3 g_5^2}{8 g_2^2 \tilde{g}_4^6} s_{10} \left( \frac{v_T}{\La_f} \right) Z  - \frac{g_5^2}{8 g_2^2 \tilde{g}_4^4 (3 \tilde{g}_3^2 + \tilde{g}_4^2)} \left(-s_5 + 2 s_{10} + \frac{g_2}{g_5} x_2\right) \left( \frac{v_T}{\La_f} \right) Z
\right.\right. \nn \\
&& \left.\left. + \frac{g_6}{6 g_2^2 \tilde{g}_3^2 \tilde{g}_4^2} \left(-\frac{3 \tilde{g}_3^2}{\tilde{g}_4^2} s_{10} -s_3-s_4 -s_5
\right) \left( \frac{v_T}{\La_f}\right) Z
- \frac{(g_1^2+3 g_2^2)}{2 g_1^2 g_2^2 \tilde{g}_4^4} (3 \tilde{g}_3^2 + \tilde{g}_4^2) s_{10} \left( \frac{v_T}{\La_f} \right)  Z
\right]
\right] \nn
\eea

\newpage
\thispagestyle{empty}
\topmargin -3 cm

\bea
\langle\varphi^S_{02}\rangle &=&   m_{SUSY} \, \left[ \frac{g_5}{4 g_2 \tilde{g}_3 \tilde{g}_4^3} Z \, + \, \left[ \frac{\tilde{g}_4}{6 g_1 \tilde{g}_3} \left(
t_6 -2 t_7 + t_8 +\frac{2 \tilde{g}_3}{\tilde{g}_4} t_9 + \frac{3 \tilde{g}_3^2}{\tilde{g}_4^2} t_{11} - \frac{2 g_1 \tilde{g}_3}{g_2 \tilde{g}_4} t_{12}
\right) \left( \frac{u}{\La_f} \right)  Y  \nn \right.\right.\\
&& \left.\left. + \frac{g_5}{2 g_2 (3 \tilde{g}_3^2 + \tilde{g}_4^2)} \left( t_6 + t_7 + t_8 + \frac{(3 \tilde{g}_3^2 -2 \tilde{g}_4^2)}{3 \tilde{g}_4^2} t_{11}
\right) \left( \frac{u}{\La_f} \right)  Y \right.\right.\nn\\
&& \left.\left. + \frac{g_0 \tilde{g}_3}{6 g_1^2 \tilde{g}_4} s_6 \left( \frac{v_T}{\La_f} \right) Y 
 + \frac{g_0 \tilde{g}_3}{4 g_1^2 \tilde{g}_4} s_8 \left( \frac{v_T}{\La_f} \right) Y
\right.\right.\nn\\
&& + \left.\left. \frac{g_0}{18 g_1^2 g_2^2} \left((-4 g_1^2 + 3 g_2^2) (s_3+s_5) -  2 (2g_1^2+3 g_2^2) s_4  - \frac{3 \tilde{g}_3^2}{\tilde{g}_4^2} (4 g_1^2-3 g_2^2)
s_{10} \right) \left( \frac{v_T}{\La_f} \right) Y \right.\right. \nn \\
&& + \left.\left. \frac{g_0 g_5^2}{18 g_2^2 \tilde{g}_4^2 (3 \tilde{g}_3^2 + \tilde{g}_4^2)} \left( -s_3 -s_4-s_5 + \frac{g_2}{g_5} x_2
-\frac{3 \tilde{g}_3^2}{\tilde{g}_4^2} s_{10} \right) \left( \frac{v_T}{\La_f} \right) Y
\right.\right. \nn \\
&& \left.\left. + \frac{1}{12 g_1^2 g_2^2} \left(\frac{1}{g_1 g_2} (-3 \mathbb{g_1} g_2^3 + 4 g_1^3 \mathbb{g_2}) s_3
 - (-3 g_2^2 + 4 g_1^2) \mathbb{s_3} \right) \left( \frac{v_T}{\La_f} \right) \right.\right. \nn \\
&& \left.\left.+ \frac{1}{6 g_1^2 g_2^2} \left(\frac{1}{g_1 g_2} (2 g_1^3 \mathbb{g_2} + 3 \mathbb{g_1} g_2^3) s_4
 - (2 g_1^2 + 3 g_2^2) \mathbb{s_4} \right) \left( \frac{v_T}{\La_f} \right)
\right.\right. \nn \\
&& \left.\left.+ \frac{1}{12 g_1^2 g_2^2} \left(\frac{1}{g_1 g_2} (4 g_1^3 \mathbb{g_2} - 3 \mathbb{g_1} g_2^3) s_5
 - (4 g_1^2 - 3 g_2^2) \mathbb{s_5} \right) \left( \frac{v_T}{\La_f} \right)
\right.\right. \nn \\
&& \left.\left. - \frac{\tilde{g}_3^2}{4 g_1^2 g_2^2 \tilde{g}_4^2} \left(\frac{1}{g_1 g_2} (3 \mathbb{g_1} g_2^3 - 4 g_1^3 \mathbb{g_2}) s_{10}
- (-4 g_1^2+3 g_2^2) \mathbb{s_{10}} \right) \left( \frac{v_T}{\La_f} \right)
\right.\right. \nn \\
&& \left.\left. - \frac{\tilde{g}_3^2 g_5}{4 g_2^3 \tilde{g}_4^4 (3 \tilde{g}_3^2 + \tilde{g}_4^2)} \left( g_2 \mathbb{g_5} - \mathbb{g_2} g_5 \right)
s_{10} \left( \frac{v_T}{\La_f} \right) + \frac{\tilde{g}_3^2 g_5^2}{4 g_2^2 \tilde{g}_4^4 (3 \tilde{g}_3^2 + \tilde{g}_4^2)}
\left( \frac{\mathbb{\tilde{g}_4}^2}{\tilde{g}_4^2} s_{10} - \mathbb{s_{10}} \right) \left( \frac{v_T}{\La_f} \right)
\right.\right. \nn \\
&&\left.\left. + \frac{g_5^2}{12 g_2^2 \tilde{g}_4^2 (3 \tilde{g}_3^2 + \tilde{g}_4^2)} \left(\left(\frac{3 \mathbb{\tilde{g}_3}^2 + \mathbb{\tilde{g}_4}^2}{3 \tilde{g}_3^2 + \tilde{g}_4^2}
s_3 - \mathbb{s_3} \right) 
 + \left(\frac{3 \mathbb{\tilde{g}_3}^2 + \mathbb{\tilde{g}_4}^2}{3 \tilde{g}_3^2 + \tilde{g}_4^2} s_4 - \mathbb{s_4} \right) \right)
\left( \frac{v_T}{\La_f} \right)
\right.\right. \nn \\
&& \left.\left.+ \frac{g_5^2}{12 g_2^2 \tilde{g}_4^2 (3 \tilde{g}_3^2 + \tilde{g}_4^2)} \left(\frac{1}{2} \left( \frac{\mathbb{\tilde{g}_4}^2}{\tilde{g}_4^2}
+ \frac{3 \mathbb{\tilde{g}_3}^2 + \mathbb{\tilde{g}_4^2}}{3 \tilde{g}_3^2 + \tilde{g}_4^2} \right) (s_5 - \frac{g_2}{g_5} x_2)  - (\mathbb{s_5} - \frac{g_2}{g_5}
\mathbb{x_2}) \right) \left( \frac{v_T}{\La_f} \right) \right.\right.\nn \\
&& \left.\left. + \frac{g_5}{12 g_2^3 \tilde{g}_4^2 (3 \tilde{g}_3^2 +\tilde{g}_4^2)} \left( \mathbb{g_2} g_5- g_2 \mathbb{g_5}\right) (s_3+s_4+s_5)
\left( \frac{v_T}{\La_f} \right) \right.\right.\nn \\
&& \left.\left.- \frac{\tilde{g}_3}{4 g_1^2 \tilde{g}_4} \left( \frac{\mathbb{g_1}}{g_1} s_6 - \mathbb{s_6} \right) \left( \frac{v_T}{\La_f} \right) 
+ \frac{3 \tilde{g}_3}{8 g_1^2 \tilde{g}_4} \left( \mathbb{s_8} - \frac{\mathbb{g_1}}{g_1} s_8\right) \left( \frac{v_T}{\La_f} \right)
\right.\right. \nn \\
&& \left.\left. + \frac{g_5}{12 g_1 g_2 \tilde{g}_3 \tilde{g}_4^3} s_1 \left( \frac{v_T}{\La_f} \right) Z 
- \frac{g_5}{4 g_2^2 \tilde{g}_4^4} s_{11} \left( \frac{v_T}{\La_f} \right) Z
\right.\right.\nn \\
&& \left.\left.- \frac{1}{12 g_1 \tilde{g}_3^2 \tilde{g}_4^2} x_{1} \left( \frac{v_T}{\La_f} \right) Z
+ \frac{1}{12 g_2 \tilde{g}_3^2 \tilde{g}_4^2} x_{3} \left( \frac{v_T}{\La_f} \right) Z  
\right.\right.\nn \\
&& \left.\left. + \frac{1}{8 g_1^2 \tilde{g}_3 \tilde{g}_4^3} ((3 \tilde{g}_3^2 +\tilde{g}_4^2) s_6 - g_1 x_2) \left( \frac{v_T}{\La_f} \right) Z
+ \frac{1}{2 g_2^2 \tilde{g}_4^2} \left(s_3 + s_4 +s_5-s_{10}\right) \left( \frac{v_T}{\La_f} \right) Z \right.\right. \nn \\
&& \left.\left. + \frac{g_5}{8 g_1 g_2 \tilde{g}_3 \tilde{g}_4^3} \left(s_2- s_3 +2 s_4 -s_5 - \frac{3 \tilde{g}_3^2}{\tilde{g}_4^2} s_{10}\right)
\left( \frac{v_T}{\La_f}\right) Z
- \frac{3 g_5^2}{8 g_2^2 \tilde{g}_4^6} s_{10} \left( \frac{v_T}{\La_f} \right) Z
\right.\right. \nn \\
&& \left.\left. + \frac{g_5^2}{24 g_2^2 \tilde{g}_3^2 \tilde{g}_4^4} \left(-s_3-s_4 -s_5 +\frac{g_2}{g_5} x_2\right) \left( \frac{v_T}{\La_f}\right) Z
\right.\right. \nn \\
&& \left.\left. - \frac{g_5^2}{8 g_2^2 \tilde{g}_4^4 (3 \tilde{g}_3^2 + \tilde{g}_4^2)} \left( -s_5 + 2 s_{10} + \frac{g_2}{g_5} x_2\right) \left( \frac{v_T}{\La_f} \right) Z \right.\right. \nn \\
&& \left.\left. + \frac{g_6}{6 g_2^2 \tilde{g}_3^2 \tilde{g}_4^2} \left(-\frac{3 \tilde{g}_3^2}{\tilde{g}_4^2} s_{10} -s_3-s_4 -s_5
\right) \left( \frac{v_T}{\La_f}\right) Z
- \frac{(2 g_1^2-3 g_2^2)}{4 g_1^2 g_2^2 \tilde{g}_4^4} (3 \tilde{g}_3^2 + \tilde{g}_4^2) s_{10} \left( \frac{v_T}{\La_f} \right)  Z \right.\right. \nn \\
&& \left.\left. - \frac{3 g_5}{8 g_1 g_2 \tilde{g}_4^4} s_8 \left(
\frac{v_T}{\La_f} \right) Z 
- \frac{3 (3\tilde{g}_3^2-\tilde{g}_4^2)}{16 g_1^2 \tilde{g}_3 \tilde{g}_4^3} s_8 \left( \frac{v_T}{\La_f} \right) Z
+ \frac{9}{8 g_1^2 \tilde{g}_4^2} \left( s_4 -s_5\right) \left( \frac{v_T}{\La_f} \right)  Z
\right]
\right] \nn
\eea

\thispagestyle{empty}

\bea
\langle\varphi^S_{03}\rangle &=& m_{SUSY} \, \left[ \frac{g_5}{4 g_2 \tilde{g}_3 \tilde{g}_4^3} Z \, + \, \left[ \frac{\tilde{g}_4}{6 g_1 \tilde{g}_3} \left(
t_6 + t_7 -2 t_8 +\frac{2 \tilde{g}_3}{\tilde{g}_4} t_9 + \frac{3 \tilde{g}_3^2}{\tilde{g}_4^2} t_{11} - \frac{2 g_1 \tilde{g}_3}{g_2 \tilde{g}_4} t_{12}
\right) \left( \frac{u}{\La_f} \right)  Y  \nn \right.\right.\\
&& \left.\left. + \frac{g_5}{2 g_2 (3 \tilde{g}_3^2 + \tilde{g}_4^2)} \left( t_6 + t_7 + t_8 + \frac{(3 \tilde{g}_3^2 -2 \tilde{g}_4^2)}{3 \tilde{g}_4^2} t_{11}
\right) \left( \frac{u}{\La_f} \right)  Y \right.\right.\nn \\
&& \left.\left.
+ \frac{g_0 \tilde{g}_3}{6 g_1^2 \tilde{g}_4} s_6 \left( \frac{v_T}{\La_f} \right) Y
 - \frac{g_0 \tilde{g}_3}{4 g_1^2 \tilde{g}_4} s_8 \left( \frac{v_T}{\La_f} \right) Y
\right.\right.\nn \\
&& + \left.\left. \frac{g_0}{18 g_1^2 g_2^2} \left(-(4 g_1^2 - 3 g_2^2) (s_3+s_4) - 2 (2 g_1^2 + 3 g_2^2) s_5 -
\frac{3 \tilde{g}_3^2}{\tilde{g}_4^2} (4 g_1^2-3 g_2^2)
s_{10} \right) \left( \frac{v_T}{\La_f} \right) Y \right.\right. \nn \\
&& + \left.\left. \frac{g_0 g_5^2}{18 g_2^2 \tilde{g}_4^2 (3 \tilde{g}_3^2 + \tilde{g}_4^2)} \left( -s_3 -s_4-s_5 + \frac{g_2}{g_5} x_2
-\frac{3 \tilde{g}_3^2}{\tilde{g}_4^2} s_{10} \right) \left( \frac{v_T}{\La_f} \right) Y
\right.\right. \nn \\
&& \left.\left. + \frac{1}{12 g_1^2 g_2^2} \left(\frac{1}{g_1 g_2} (-3 \mathbb{g_1} g_2^3 + 4 g_1^3 \mathbb{g_2}) s_3
 - (4 g_1^2 - 3 g_2^2) \mathbb{s_3} \right) \left( \frac{v_T}{\La_f} \right) \right.\right. \nn \\
&& \left.\left.+ \frac{1}{12 g_1^2 g_2^2} \left(\frac{1}{g_1 g_2} (4 g_1^3 \mathbb{g_2} - 3 \mathbb{g_1} g_2^3) s_4
 - (4 g_1^2 - 3 g_2^2) \mathbb{s_4} \right) \left( \frac{v_T}{\La_f} \right)
\right.\right. \nn \\
&& \left.\left.+ \frac{1}{6 g_1^2 g_2^2} \left(\frac{1}{g_1 g_2} (2 g_1^3 \mathbb{g_2} + 3 \mathbb{g_1} g_2^3) s_5
 - (2 g_1^2 + 3 g_2^2) \mathbb{s_5} \right) \left( \frac{v_T}{\La_f} \right)
\right.\right. \nn \\
&& \left.\left. + \frac{\tilde{g}_3^2}{4 g_1^2 g_2^2 \tilde{g}_4^2} \left(\frac{1}{g_1 g_2} (-3 \mathbb{g_1} g_2^3 + 4 g_1^3 \mathbb{g_2}) s_{10}
- (4 g_1^2-3 g_2^2) \mathbb{s_{10}} \right) \left( \frac{v_T}{\La_f} \right)
\right.\right. \nn \\
&& \left.\left. - \frac{\tilde{g}_3^2 g_5}{4 g_2^3 \tilde{g}_4^4 (3 \tilde{g}_3^2 + \tilde{g}_4^2)} \left( g_2 \mathbb{g_5} - \mathbb{g_2} g_5 \right)
s_{10} \left( \frac{v_T}{\La_f} \right) + \frac{\tilde{g}_3^2 g_5^2}{4 g_2^2 \tilde{g}_4^4 (3 \tilde{g}_3^2 + \tilde{g}_4^2)}
\left( \frac{\mathbb{\tilde{g}_4}^2}{\tilde{g}_4^2} s_{10} - \mathbb{s_{10}} \right) \left( \frac{v_T}{\La_f} \right)
\right.\right. \nn \\
&&\left.\left. + \frac{g_5^2}{12 g_2^2 \tilde{g}_4^2 (3 \tilde{g}_3^2 + \tilde{g}_4^2)} \left(\left(\frac{3 \mathbb{\tilde{g}_3}^2 + \mathbb{\tilde{g}_4}^2}{3 \tilde{g}_3^2 + \tilde{g}_4^2}
s_3 - \mathbb{s_3} \right) 
+ \left(\frac{3 \mathbb{\tilde{g}_3}^2 + \mathbb{\tilde{g}_4}^2}{3 \tilde{g}_3^2
+ \tilde{g}_4^2} s_4 - \mathbb{s_4} \right) \right) \left( \frac{v_T}{\La_f} \right)
\right.\right. \nn \\
&& \left.\left.+ \frac{g_5^2}{12 g_2^2 \tilde{g}_4^2 (3 \tilde{g}_3^2 + \tilde{g}_4^2)} \left(\frac{1}{2} \left( \frac{\mathbb{\tilde{g}_4}^2}{\tilde{g}_4^2}
+ \frac{3 \mathbb{\tilde{g}_3}^2 + \mathbb{\tilde{g}_4^2}}{3 \tilde{g}_3^2 + \tilde{g}_4^2} \right) (s_5 - \frac{g_2}{g_5} x_2)  - (\mathbb{s_5} - \frac{g_2}{g_5}
\mathbb{x_2}) \right) \left( \frac{v_T}{\La_f} \right) \right.\right.\nn \\
&& \left.\left. + \frac{g_5}{12 g_2^3 \tilde{g}_4^2 (3 \tilde{g}_3^2 +\tilde{g}_4^2)} \left( \mathbb{g_2} g_5- g_2 \mathbb{g_5}\right) (s_3+s_4+s_5)
\left( \frac{v_T}{\La_f} \right)\right.\right. \nn \\
&& \left.\left.
- \frac{\tilde{g}_3}{4 g_1^2 \tilde{g}_4} \left( \frac{\mathbb{g_1}}{g_1} s_6 - \mathbb{s_6} \right) \left( \frac{v_T}{\La_f} \right) 
- \frac{3 \tilde{g}_3}{8 g_1^2 \tilde{g}_4} \left( \mathbb{s_8} - \frac{\mathbb{g_1}}{g_1} s_8\right) \left( \frac{v_T}{\La_f} \right)
\right.\right. \nn \\
&& \left.\left. + \frac{g_5}{12 g_1 g_2 \tilde{g}_3 \tilde{g}_4^3} s_1 \left( \frac{v_T}{\La_f} \right) Z 
- \frac{g_5}{4 g_2^2 \tilde{g}_4^4} s_{11} \left( \frac{v_T}{\La_f} \right) Z
\right.\right.\nn \\
&& \left.\left.- \frac{1}{12 g_1 \tilde{g}_3^2 \tilde{g}_4^2} x_{1} \left( \frac{v_T}{\La_f} \right) Z
+ \frac{1}{12 g_2 \tilde{g}_3^2 \tilde{g}_4^2} x_{3} \left( \frac{v_T}{\La_f} \right) Z
\right.\right.\nn \\
&& \left.\left. - \frac{1}{8 g_1^2 \tilde{g}_3 \tilde{g}_4^3} (-(3 \tilde{g}_3^2 + \tilde{g}_4^2) s_6 + g_1 x_2) \left( \frac{v_T}{\La_f} \right) Z
+ \frac{1}{2 g_2^2 \tilde{g}_4^2} \left(s_3 + s_4 +s_5 -s_{10}\right) \left( \frac{v_T}{\La_f} \right) Z \right.\right. \nn \\
&& \left.\left. + \frac{g_5}{8 g_1 g_2 \tilde{g}_3 \tilde{g}_4^3} \left(-s_2-s_3 -s_4 + 2 s_5 - \frac{3 \tilde{g}_3^2}{\tilde{g}_4^2} s_{10}\right)
\left( \frac{v_T}{\La_f}\right) Z
- \frac{3 g_5^2}{8 g_2^2 \tilde{g}_4^6} s_{10} \left( \frac{v_T}{\La_f} \right) Z
\right.\right. \nn \\
&& \left.\left. + \frac{g_5^2}{24 g_2^2 \tilde{g}_3^2 \tilde{g}_4^4} \left(-s_3-s_4 -s_5 +\frac{g_2}{g_5} x_2\right) \left( \frac{v_T}{\La_f}\right) Z \right.\right.
\nn \\
&& \left.\left. - \frac{g_5^2}{8 g_2^2 \tilde{g}_4^4 (3 \tilde{g}_3^2 + \tilde{g}_4^2)} \left(-s_5 +2 s_{10} +\frac{g_2}{g_5} x_2\right)
\left( \frac{v_T}{\La_f} \right) Z \right.\right. \nn \\
&& \left.\left. + \frac{g_6}{6 g_2^2 \tilde{g}_3^2 \tilde{g}_4^2} \left(-\frac{3 \tilde{g}_3^2}{\tilde{g}_4^2} s_{10} -s_3-s_4 -s_5
\right) \left( \frac{v_T}{\La_f}\right) Z
- \frac{(2 g_1^2-3 g_2^2)}{4 g_1^2 g_2^2 \tilde{g}_4^4} (3 \tilde{g}_3^2 + \tilde{g}_4^2) s_{10} \left( \frac{v_T}{\La_f} \right)  Z
\right.\right.\nn \\
&& \left.\left. + \frac{3 g_5}{8 g_1 g_2 \tilde{g}_4^4} s_8 \left(
\frac{v_T}{\La_f} \right) Z  + \frac{3 (3\tilde{g}_3^2-\tilde{g}_4^2)}{16 g_1^2 \tilde{g}_3 \tilde{g}_4^3} s_8 \left( \frac{v_T}{\La_f} \right) Z
- \frac{9}{8 g_1^2 \tilde{g}_4^2}  \left( s_4 -s_5\right) \left( \frac{v_T}{\La_f} \right) Z
\right]
\right] \nn
\eea


\newpage
\topmargin -2 cm

\end{document}